\documentclass[eqsecnum,preprint,flushrt]{aastex}
\usepackage{natbib}
\usepackage{lscape}
\usepackage{gensymb}
\usepackage{url}
\usepackage{graphicx}
\usepackage{multicol}
\usepackage[title]{appendix}

\usepackage{amsmath, amsthm, amssymb }

\begin{document}

\title{Magnetic-Island Contraction and Particle Acceleration \\ in Simulated Eruptive Solar Flares}

\author{S.\ E.\ Guidoni}
\affil{The Catholic University of America, 620 Michigan Avenue Northeast, Washington, DC 20064}
\affil{Heliophysics Science Division, NASA Goddard Space Flight Center, Greenbelt, MD 20771}
\author{C.\ R.\ DeVore and J.\ T.\ Karpen}
\affil{Heliophysics Science Division, NASA Goddard Space Flight Center, Greenbelt, MD 20771}
\author{B.\ J.\ Lynch}
\affil{Space Sciences Laboratory, University of California, Berkeley, CA 94720}

\email{silvina.e.guidoni@nasa.gov}

\begin{abstract}

The mechanism that accelerates particles to the energies required to produce the observed high-energy impulsive emission in solar flares is not well understood. Drake et al. (2006) proposed a mechanism for accelerating electrons in contracting magnetic islands formed by kinetic reconnection in multi-layered current sheets. We apply these ideas to sunward-moving flux ropes (2.5D magnetic islands) formed during fast reconnection in a simulated eruptive flare. A simple analytic model is used to calculate the energy gain of particles orbiting the field lines of the contracting magnetic islands in our ultrahigh-resolution 2.5D numerical simulation. We find that the estimated energy gains in a single island range up to a factor of five. This is higher than that found by Drake et al. for islands in the terrestrial magnetosphere and at the heliopause, due to strong plasma compression that occurs at the flare current sheet. In order to increase their energy by two orders of magnitude and plausibly account for the observed high-energy flare emission, the electrons must visit multiple contracting islands. This mechanism should produce sporadic emission because island formation is intermittent. Moreover, a large number of particles could be accelerated in each magnetohydrodynamic-scale island, which may explain the inferred rates of energetic-electron production in flares. We conclude that island contraction in the flare current sheet is a promising candidate for electron acceleration in solar eruptions.

\end{abstract}


%

\section{Introduction}

The hard X-ray (HXR) and microwave radiation from solar flares is generally accepted to be emitted by nonthermal particles accelerated during the release of magnetic energy that powers the flare, when they interact with the ambient atmosphere. HXR emission is bremsstrahlung from accelerated electrons that lose their energy through collisions (\citealt{Holman_2011}) and microwave emission is gyrosynchrotron from accelerated electrons that spiral the coronal magnetic field (\citealt{White_2011}). On the other hand, extreme ultraviolet (EUV), coronal soft X-ray, and other secondary emissions result from the thermalization of flare energy in the ambient atmosphere \citep[see reviews by][]{Benz_2008, Fletcher_2011}. Therefore HXR and microwave flare observations, particularly during the initial impulsive phase, offer valuable clues about the physics of the particle-acceleration process \citep[e.g.,][]{Holman_2011}. Decades of study have established that these emissions often are highly variable in time \citep[e.g.,][]{Parks_1969, Chiu_1970, Kane_1983, Kiplinger_1983, Kiplinger_1984, Machado_1993, Aschwanden_1995_I} and originate at both the tops and footpoints of coronal loops \citep{Marsh_1980, Hoyng_1981, Masuda_1994, Reznikova_2009, Caspi_2010, Ishikawa_2011}. More recently, a second coronal source has been discovered above the loop-top coronal source (\citealt{Sui_2003, Sui_2004, Pick_2005, Liu_2008, Su_2012, Liu_2013}).  

During flares, HXR pulses lasting milliseconds to several seconds, superposed on more slowly varying background emission, have been observed by NASA's Reuven Ramaty High Energy Solar Spectroscopic Imager \citep[{\em RHESSI}; e.g.,][]{Holman_2011, Inglis_2012}. These pulses are most frequently seen during the early flare impulsive phase, but occasionally are seen during the late decay phase. Correlated temporal intermittencies of impulsive emissions in microwave and HXR indicate a common, underlying, intermittent particle-acceleration mechanism (\citealt{Aschwanden_1995_I, Inglis_2009, Nakariakov_2010}). Short trains of multiple pulses with varying periods and amplitudes, denoted ``quasi-periodic pulsations,'' have been reported (\citealt{Nakariakov_2007, Inglis_2008, Inglis_2009, Nakariakov_2009, Inglis_2012, Inglis_2013}), although whether they are truly periodic remains a matter of debate \citep{Vaughan_2010, Gruber_2011, Inglis_2015}. 

In the standard model for eruptive two-ribbon solar flares \citep[e.g.,][]{Carmichael_1964, Sturrock_1966, Hirayama_1974, Kopp_1976}, the outward stretching of the magnetic structure that ultimately comprises the coronal mass ejection (CME) forms a vertical electric current sheet (CS) in the wake of the CME. Reconnection across this sheet \citep[e.g.,][]{Sweet_1958, Parker_1963, Petschek_1964} drives Alfv\'enic outflows and shock waves, forms the flare arcade loops and the ejected CME flux rope, and relaxes the stressed system toward a lower-energy state. Although the original descriptions of the model posited a single reconnection site, numerous high-resolution studies of both kinetic and magnetofluid systems at high Lundquist number demonstrate that CSs develop multiple reconnection sites with strong spatial (``patchy'') and temporal (``bursty'') variability \citep[e.g.,][]{Daughton_2006, Daughton_2014, Drake_2006_I, Loureiro_2007, Samtaney_2009, Fermo_2010, Uzdensky_2010, Mei_2012, Huang_2012, Karpen_2012, Cassak_2013,  Shen_2013}. The plasmoids (also called magnetic islands in two dimensions) formed by reconnection can fragment, coalesce, or remain intact, as determined in part by the boundary conditions, the initial condition (e.g., isolated sheets vs.\ multiple interacting sheets), and whether the reconnection is driven or spontaneous. Several investigations relate solar flare pulsations to plasmoid formation by reconnection (see, for example, \citealt{Kliem_2000, Karlicky_2004, Karlicky_2007, Barta_2008}). Dynamic plasmoid structures have been observed in a few well-resolved, large-scale simulations of eruptive solar flares (e.g., \citealt{Karpen_2012, Guo_2013}). Spatial intermittencies, in the form of bright dynamic blobs, have been observed in high-resolution EUV data from NASA's Solar and Heliospheric Observatory ({\em SOHO}), Solar TErrestrial RElations Observatory ({\em STEREO}), and Solar Dynamics Observatory ({\em SDO})(\citealt{Ohyama_1998, Ko_2003, Lin_2005, Riley_2007, Reeves_2008, Lin_2008, Ciaravella_2008, Takasao_2012, Kumar_2013, Liu_2013}). These discrete features generally emanate from positions above the flare loops, and often appear to be spaced along a common line like beads on a wire.

Processes proposed to accelerate flare electrons include the reconnection electric field itself, induced electric fields associated with the reconfiguration of the reconnected magnetic field, stochastic scattering by turbulent magnetohydrodynamic (MHD) waves excited in the reconnection region, and parallel electric fields associated with kinetic-scale Alfv\'en-wave cascades downstream from the flare site \citep[see the review by][]{Zharkova_2011}. The second process includes the acceleration of particles during magnetic-island contraction, discovered via kinetic simulations  \citep{Drake_2005, Drake_2006, Drake_2006_I, Drake_2010, Drake_2013, Fermo_2010}. In essence, the particles gain energy by reflecting off converging magnetic mirrors -- in this case, the reconfigured field lines of the contracting islands -- as proposed originally by \citet{Fermi_1949} in his model for cosmic-ray acceleration. Whereas a fully self-consistent treatment of the problem requires calculating individual particle orbits and feedback from the particles to the electromagnetic fields, \citet{Drake_2010} derived some essential features of the acceleration process using a kinetic approach based on adiabatic invariants of the particle motion. Here, we use an analytic method extending the ideas presented in that paper.

Although the partitioning of flare energy among kinetic energy, thermal energy, radiation, and accelerated particles is not well established (\citealt{Emslie_2004, Emslie_2005, Raymond_2008}), a significant fraction of the released magnetic free energy must be used in electron acceleration to explain the observed X-ray intensities (\citealt{Miller_1997}).  A large number of particles are accelerated during the brief discrete energy releases: for example, \citet{Kiplinger_1983} estimated that more than $10^{34}$ electrons were accelerated to energies greater than $20$ keV in one  $400$ ms spike. A major flare can accelerate more than $10^{36}$ electrons s$^{-1}$ (\citealt{Hoyng_1976, Holman_2003}). Any proposed acceleration mechanism must be highly efficient to provide such a large number of accelerated particles (known as the ``number problem''; \citealt{Brown_1977}). Given the low density of the preflare corona and the required electron numbers, it is highly unlikely that most or all of the electron acceleration can occur within the small diffusion region in reconnection sites. More likely, the major acceleration occurs at the reconnection outflows along the reconnecting CS, as observations of double coronal sources seem to indicate (\citealt{Liu_2008, Liu_2013}). 

The objective of our paper is to apply the mechanism of charged-particle acceleration by magnetic-island contraction to the results of a large-scale MHD simulation of a solar eruption, and to assess the implications of the model for understanding observed high-energy flare emissions. For analytical simplicity, and in order to maximize the numerical resolution of our simulations of eruptive flares (EFs), we restrict our attention to an axisymmetric configuration with two-dimensional spatial variations but fully three-dimensional vector fields (a so-called 2.5D system). Our analytic method of calculating the particle energy gain in contracting flux ropes is presented in detail in Appendices \ref{sec:appen_A}-\ref{sec:appen_C}; its key findings are summarized in \S \ref{sec:analytical_A}. In \S \ref{sec:breakout}, we describe the main features of our simulated EF initiated by the magnetic breakout mechanism \citep{Antiochos_1999_I}, in which we extend previous work \citep[][hereafter KAD12]{Karpen_2012} to even finer grids that capture in greater detail the island evolution. Our analysis of island evolution and the associated energy amplification of particles in the flare CS are presented in \S \ref{sec:results}. The implications of our results for the coronal origin of electrons accelerated in solar flares and suggestions for further work are discussed in \S \ref{sec:discussion}. 

\section{Particle Acceleration via Island Contraction}
\label{sec:analytical_A}

The acceleration of particles to high energies in contracting magnetic islands has been studied extensively by Drake and his collaborators \citep{Drake_2005, Drake_2006, Drake_2006_I, Drake_2010, Drake_2013, Fermo_2010} using particle-in-cell kinetic simulations \citep[see also][]{Zank_2014}. Drake et al.\ assumed that many parallel CSs are present, as can occur if an initially extended single sheet bends and folds upon itself. Under these conditions, particles undergo multiple episodes of acceleration as they become energized within one contracting island, then drift or are scattered to another, where the process begins anew. In solar EFs, on the other hand, the energization process is expected to occur in islands within the single, dynamically evolving flare sheet formed in the wake of the CME.

\begin{figure}[h!]
   \centering
   \begin{center}$
      \begin{array}{cc}
        \resizebox{6.0in}{!}{\includegraphics{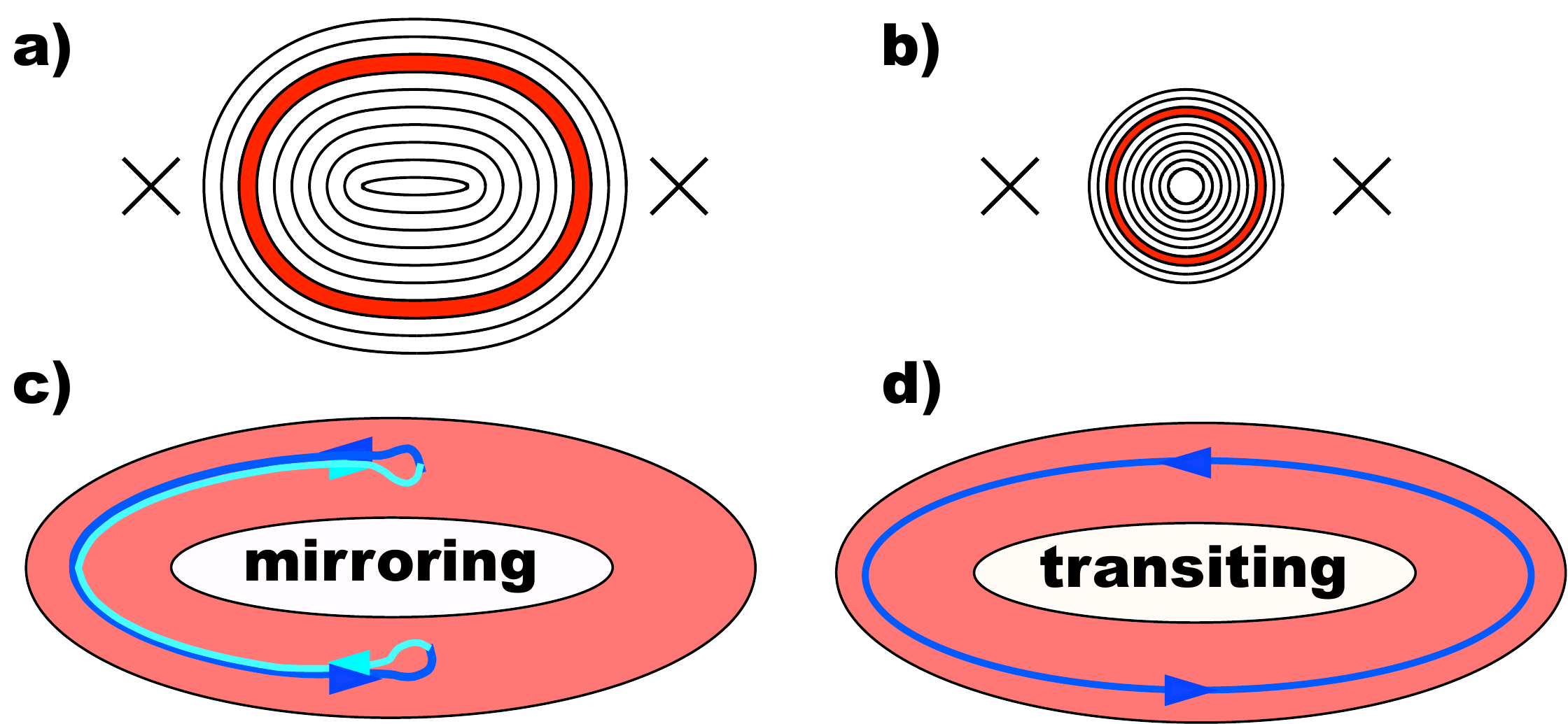}}
      \end{array}$
   \end{center}
\caption {Top: schematic diagram of contraction of a 2D magnetic island. a) Early state of a two-dimensional elongated (elliptical) magnetic island. b) Late state of the island after contraction to a perimeter-minimizing (circular) shape. The red area shows a generic flux region, before and after island contraction. Bottom: sketch of particle orbits along field lines within the red region of a) and b) (the gyromotion around field lines is not shown).  c) Particle mirrors at turning points of its orbit in the contracting island. d) Particle transits completely around the contracting island.}
\label{fig:island_cartoon}
\end{figure}

The time-varying geometry of such a contracting island is shown schematically in Figure \ref{fig:island_cartoon}, from (a) an initial state with an elongated configuration to (b) a final state with a more circular structure. The curves represent magnetic flux surfaces; the red region is a narrow volume between two such surfaces. After an island is formed during reconnection, the retraction of the field lines from the reconnection sites (represented by X symbols to the left and right of the island), due to the Lorentz force, propels the field lines toward the center, reshaping the island as shown. In drawing this evolution, we have not assumed that the area between every pair of flux surfaces is preserved; that is, we do not assume that plasma is incompressible. This is a significant difference between the flare environment and the kinetic applications: the plasma evolution in flares is strongly compressible, whereas the evolution simulated by \citet{Fermo_2010} and analyzed by \citet{Drake_2010} is essentially incompressible.

We developed an analytic method to estimate the energy gained by particles in orbits around 2.5D magnetic field lines in the general case where the island has an out-of-plane (toroidal) magnetic field component. The detailed calculations are presented in Appendix \ref{sec:appen_A}, so in this section we present the key results. We assumed that the field lines are part of a flux rope (Fig.\ \ref{fig:island_cartoon} sketches their projections in the plane perpendicular to the flux rope axis) and the flux rope is translationally symmetric in the out-of-plane direction. We considered two-dimensional spatial variations, but fully 3D particle velocities and magnetic fields. Our method is based on the conservation of the magnetic moment and parallel action of particles whose orbits are completed much more quickly than the magnetic field changes in time.

Once a flux surface of the flux rope is created (after two reconnection episodes), particles magnetically linked to that surface will orbit along it. We use subscript {\it i} to denote the initial values of quantities at the time, $t_{i}$, that a flux surface is created. The orbiting particles have a range of speeds $V$ and pitch angles $\theta$ (angle between the particle velocity and the local magnetic field). The type of motion that a particle executes depends on the magnetic field strength $B$ along the field line and the particle's pitch angle. For simplicity, we assume that the island is symmetric left/right and up/down; therefore the field strength of a field line inside the island possesses two equal minima $B_{1}$ (left and right in Fig.\ \ref{fig:island_cartoon}, located near the magnetic X points) and two equal maxima $B_{2}$ (located near the arrow heads in Fig.\ \ref{fig:island_cartoon}). $B$ is parameterized by the arc-length coordinate $l$ along the field line as
\begin{equation}
   \label{eq:B_prof_res}
    B = B_{1} + \left( B_{2}-B_{1} \right) \sin^{2} \left( 2 \pi \frac{l}{L} \right),
\end{equation}
which is a reasonable representation for our simulated islands (see \S \ref{sec:island_evol}). The length of one turn of the flux rope is $L$. Note that $B_{1}$, $B_{2}$, and $L$ may be time dependent and vary from one flux surface to another inside the flux rope, and $t_{i}$ will be different for each flux surface of the island (those closer to the axis of the flux rope are created earlier than those farther away). The approach used in Appendix \ref{sec:appen_A} to derive the energy amplification factors of particles orbiting these field lines is quite general, but this particular profile for $B$ allows the results to be expressed analytically in closed form.

We identify two types of particle orbits: \textit{mirroring} (Fig.\ \ref{fig:island_cartoon}c) and \textit{transiting} (Fig.\ \ref{fig:island_cartoon}d); the ever-present gyro-rotation about the magnetic field is not shown in Figure \ref{fig:island_cartoon}. Which type of orbit a particle executes is determined by the particle's pitch angle, $\theta_{1}$, when the particle is at the minimum of the field, $B=B_{1}$, and by the mirror ratio of the field line, $R_{m} = B_{2}/B_{1}$ (see Appendix \ref{sec:appen_A}). If $\sin\theta_{1} \geq 1/\sqrt{R_{m}}$, the particle mirrors at or before the maximum of the field (Fig.\ \ref{fig:island_cartoon}c). On the other hand, if $\sin\theta_{1} < 1/\sqrt{R_{m}}$, the particle traverses the entire island and its velocity parallel to the field never changes sign (Fig.\ \ref{fig:island_cartoon}d). Transiting particles move from one turn of the flux rope to the next, whereas mirroring particles are trapped in the original turn of the flux rope where they were injected. \citet{Fermi_1949} described the energy gain of particles reflected by moving magnetic mirrors as being due to ``type A'' and ``type B'' acceleration for mirroring and transiting particles, respectively. 

Over a single turn of a particle's orbit, the energy change is very small. We characterize the initial energy of the particle by its energy when the particle is at the minimum of the field, $E_{1i} = m V_{1i}^{2} / 2$. After many turns on its orbit, the particle's final energy, $E_{1} = m V_{1}^{2}/2$, will have increased, as the magnetic field slowly changed. In Appendix \ref{sec:appen_A}, we show that the ratio of these two energies can be expressed as follows: 
  \begin{align}
      \label{eq:en_gain_results}
      \mathcal{E}  & =  \frac{E_{1}}{E_{1i}} = \sin^2 \theta_{1i}    \hbox{ } \mathcal{E}_{\perp}  +  \cos^2 \theta_{1i}   \hbox{ }  \mathcal{E}_{\parallel}.
  \end{align}
$\theta_{1i}$ is the pitch angle of the particle at $t_{i}$, and $\mathcal{E}_{\perp}$ and $\mathcal{E}_{\parallel}$ are the energy gains in the directionperpendicular and parallel to the magnetic field, respectively:
 \begin{equation}
       \label{eq:E_perp_results}
      \mathcal{E}_{\perp}  =    \frac{V^2_{\perp 1}}{V^2_{\perp 1i}}  =  \frac{B_{1}}{B_{1i}},
 \end{equation}
  \begin{equation}
      \label{eq:E_paral_results}
      \mathcal{E}_{\parallel}  =  \frac{V^2_{\parallel 1}}{V^2_{\parallel 1i}} =  \left( \frac{B_{1}}{B_{1i}} \right) \left( \frac{\tan^2 \theta_{1i}}{\tan^2 \theta_{1}}\right).
 \end{equation}
Here $V_{\perp}$ and $V_{\parallel}$ are the components of particle velocity perpendicular and parallel to the magnetic field, respectively. Therefore, the particle energy gain is a function of the field strength and the particle's initial and final pitch angles at the minimum of the field, and it is independent of the particle's initial speed. In all cases, if the strength of the field increases as the island evolves, the perpendicular energy increases. The parallel energy gain depends upon the evolution of the pitch angle relative to the field strength. The energy multipliers $\mathcal{E}_{\perp}$ and $\mathcal{E}_{\parallel}$ are different, showing that an initially isotropic energy becomes increasingly anisotropic as the islands evolve. 

In Appendix \ref{sec:appen_A}, we derive an analytic expression for the final pitch angle $\theta_{1}$ as a function of the initial angle $\theta_{1i}$. In practice, $\theta_{1}$ is calculated by numerically solving transcendental equations that are functions of the field line properties: the minimum field strength, the mirror ratio, and the length of one turn of the field line. Because our simulations described in \S \ref{sec:breakout} provide us with these field line properties, we can compute the energy gain for particles at all initial pitch angles as our simulated islands evolve (\S \ref{sec:part_en_gain}). The general results in Appendix \ref{sec:appen_A} simplify to algebraic expressions in the limiting case of uniform field strength, $B_{2} = B_{1}$, in which there are no mirroring particles. For this case (Appendix \ref{sec:appen_B}), we show that the toroidal (out of plane) field has the net effect of moderating the impact of changes in the island perimeter on energy gain, and that plasma compression is also relevant for particle energy change. In Appendix \ref{sec:appen_B}, we also compare our results with previous analytical results from \citet{Drake_2010}. In Appendix \ref{sec:appen_C}, we estimate the range of validity of our analysis: it is appropriate for a significant fraction of the coronal electron population, though only for protons and ions that already are strongly superthermal. Therefore, our approach is adequate for a first assessment of the contribution of magnetic-island contraction to electron acceleration in solar flares.

\begin{figure}[h!]
   \centering
   \begin{center}$
      \begin{array}{cc}
        \resizebox{ 6.0in}{!}{\includegraphics{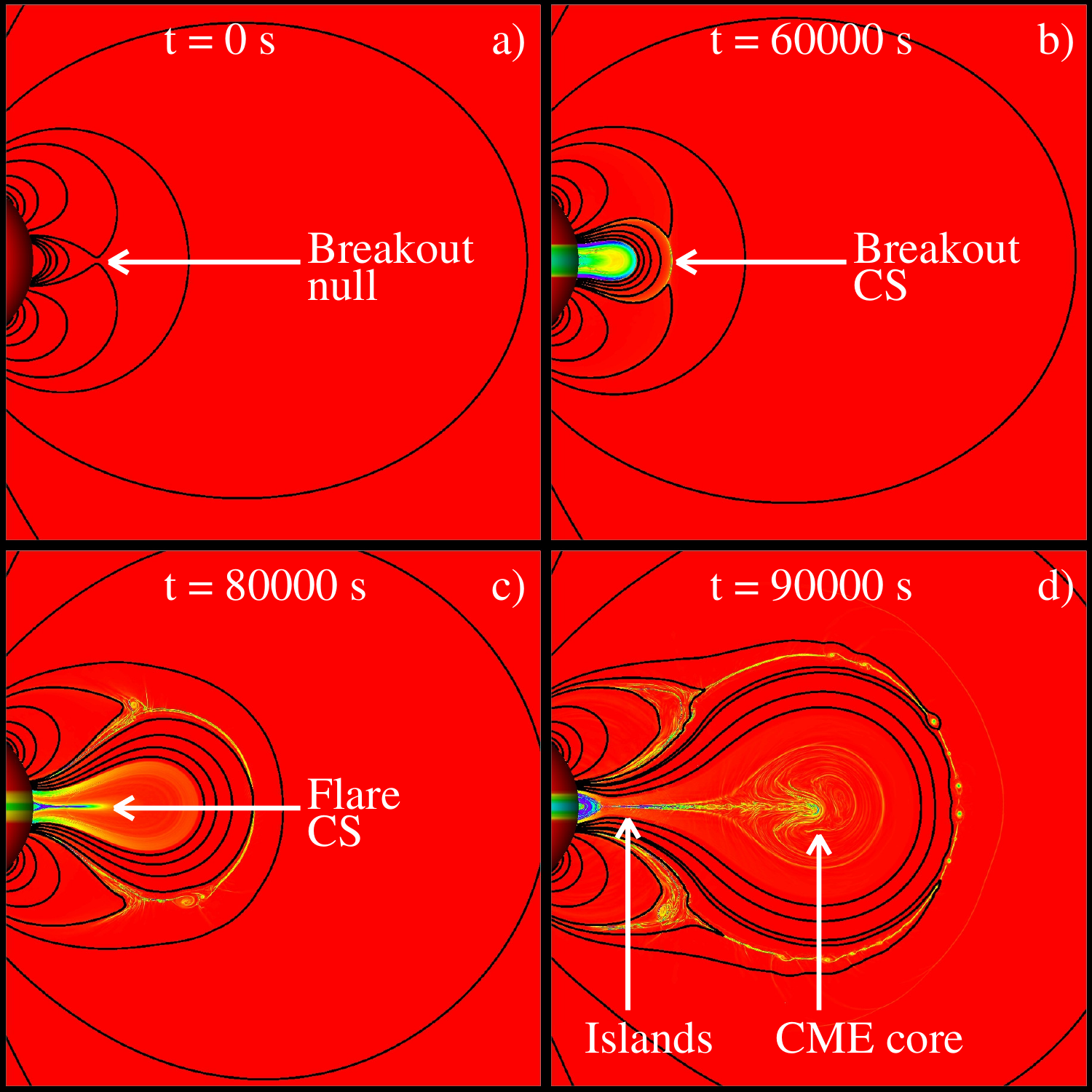} }
      \end{array}$
   \end{center}
\caption {Key structures of the breakout model at selected times of the simulation. The spherical surface is at $1$ solar radius ($R_{s}$) and shows photospheric $B_{\phi}$, with shading from red ($0$ G) to magenta ($3$ G). Normalized current density magnitude, $R_{s} J/c$ ($c$ is the speed of light), in the $r - \theta$ plane is shown with shading from red ($0$ G) to magenta (saturated at $1.5$ G).  Magnetic field lines (black) are drawn from the same set of footpoints for each time. a) $t = 0$ s (initial potential state);  b) $t = 60,000$ s (before onset of breakout reconnection); c) $t = 80,000$ s (after breakout reconnection onset and before flare reconnection); d) $t = 90,000$ s (after flare reconnection onset). The full temporal evolution is shown in the animation available in the online version (f2\_movie.mpg).}
\label{fig:breakout}
\end{figure}

\section{Simulated Breakout CME/EF}
\label{sec:breakout}

We investigate the generation and evolution of islands in a flare CS by simulating a fast CME/EF according to the well-studied breakout paradigm (\citealt{Antiochos_1998, Antiochos_1999_I, MacNeice_2004, DeVore_2005, DeVore_2008, Lynch_2008, Lynch_2009, Karpen_2012, Masson_2013}). This model has been described thoroughly in previous papers, so we only summarize the fundamental aspects here.  Key structures and evolutionary stages in the model are illustrated in Figure \ref{fig:breakout}. A fast breakout CME/EF requires three essential ingredients: two or more flux systems in the corona, yielding at least one magnetic null point; slow accumulation of magnetic free energy, through either flux emergence or (as in our simulations) footpoint motions; and a small but finite resistivity that allows reconnection to occur.  Given these three elements, eruption is inevitable once the critical amount of free energy has accumulated in the system (see KAD12). Reconnection plays two key roles in this scenario: breakout reconnection at the preexisting coronal null removes overlying flux that restrains the stressed core field from erupting, and flare reconnection partially detaches the erupting core, thus separating the flux rope that comprises the CME from the arcade of flare loops below. 

Our previous studies demonstrated the viability of this model in 2.5D and 3D, with and without the presence of a solar wind. A recent investigation (KAD12) found that a resistive instability in the flare CS is responsible for triggering the explosive eruption, thus identifying a physical basis for the strong observed connection between fast CMEs and two-ribbon flares. This conclusion followed from our analysis of an ultra-high-resolution, 2.5D MHD simulation of a breakout eruption performed with the Adaptively Refined MHD Solver (ARMS). ARMS employs a finite-volume representation of the plasma and magnetic field, high-fidelity Flux-Corrected Transport techniques \citep{DeVore_1991}, and adaptive mesh refinement managed by the PARAMESH toolkit \citep{MacNeice_2000}. As in all finite-volume solutions to the MHD equations, numerical resistivity permits magnetic reconnection to occur at electric CSs that thin down to the finite scale of the simulation grid. Adaptivity criteria tailored to the developing CSs allowed us to achieve the highest overall spatial resolution to date in a CME/EF simulation. Consequently, the high-cadence KAD12 simulation followed the self-consistent formation of the breakout and flare CSs, and tracked the dynamic life cycle of numerous magnetic islands in both CSs. The null-finding capability implemented by KAD12 enables individual nulls to be identified and characterized as X- and O-types (\citealt{Parnell_1996}).

We found that still higher resolution is required to analyze the changing shapes of the islands in the flare sheet, in order to estimate the resulting particle energization described in \S \ref{sec:results}. Therefore, in the present work, we use the same initial and boundary conditions as KAD12 but increase the maximum resolution from $6$ to $8$ levels of refinement. A detailed description of the refinement criteria is given in KAD12. The minimum distance between grid points $d_{g}$ achieved in our simulated flare CS is $\approx 0.001 R_{s} = 700$ km, comparable to the resolution of {\em SDO}'s Atmospheric Imaging Array. 

The initial potential-field configuration of our simulation represents the Sun's global dipole field plus a large equatorial active region (see Fig.\ \ref{fig:breakout}a), the simplest configuration that meets the first (multipolar) requirement for the breakout mechanism. As in KAD12, the maximum initial field strength at the solar poles is $10$ G. ARMS solved the ideal 2.5D MHD equations in spherical coordinates $(r,\theta,\phi)$; see KAD12 for a full description of the equations. Figure \ref{fig:breakout} and the accompanying animation (available in the online journal) illustrate the evolution of the system from its initial potential state through eruption and flaring. As in our previous breakout studies, free energy was added slowly to the system by driving antiparallel footpoint motions in a narrow band centered on the equatorial polarity inversion line (PIL). The flow velocities are well below the Alfv\'en and sound speeds at the lower boundary. The resulting excess magnetic pressure compresses the initial coronal null point (Fig.\ \ref{fig:breakout}a) into a breakout CS (Fig.\ \ref{fig:breakout}b) that eventually thins down to the grid scale. Thereafter, reconnection steadily removes overlying flux enabling the core to expand. Behind the expanding core, the flare CS forms, lengthens, and thins down to the grid scale (Fig.\ \ref{fig:breakout}c); it extends to the top of the flare arcade and is roughly aligned with the equatorial radial direction. When fast reconnection starts at the flare CS, much of the magnetic free energy is released rapidly as the flux rope detaches and strongly accelerates outward. Islands formed by the flare reconnection move along this CS (Fig.\ \ref{fig:breakout}d). 

\section{Analysis and Results}
\label{sec:results}

\subsection{Flare Current Sheet Nulls}
\label{sec:flare_nulls}

\begin{figure}[h!]
   \centering
   \begin{center}$
      \begin{array}{cc}
          \resizebox{6.0in}{!}{\includegraphics{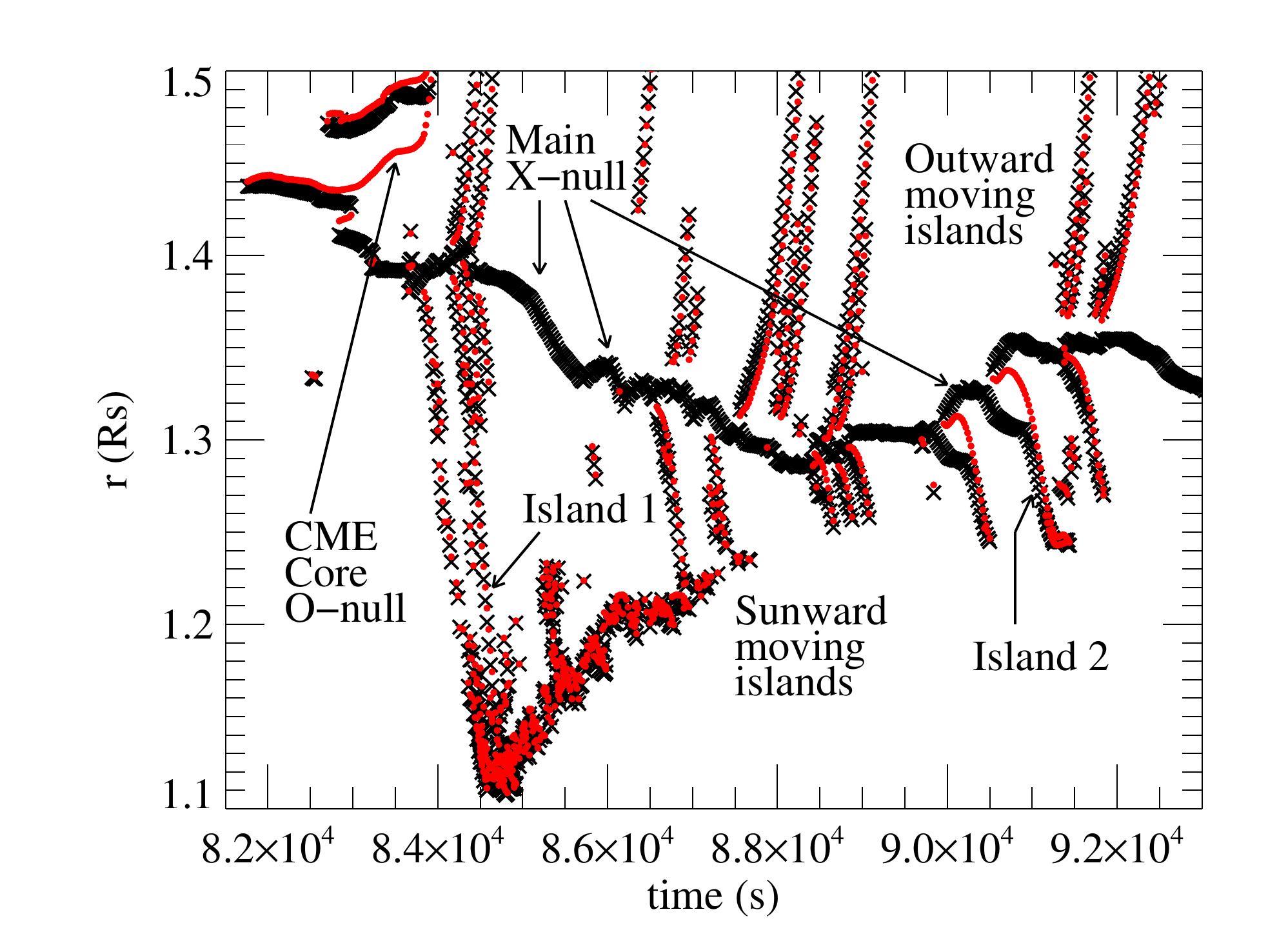}}
     \end{array}$
   \end{center}
\caption {Radial position of X- (black crosses) and O- (red circles) nulls as a function of time. The evolution of Island 1 and Island 2 is discussed in \S \ref{sec:island_evol}.}
 \label{fig:All_nulls} 
\end{figure}

Reconnection in the flare CS begins at time $t\approx81760$ s, when a single X-null forms at $r = 1.4375$ $R_{s}$. Note that our ``nulls'' are locations where the 2D poloidal magnetic field (components in the $r-\theta$ plane) is zero, although the toroidal field (in the $\phi$ direction) is generally non-zero. Only a few transient, slow-moving nulls form in the flare sheet at first, but the reconnection quickly transitions at $t \approx 84000$ s to produce fast Alfv\'enic jets and a series of sunward- and outward-moving islands. Each island is an O-null formed by reconnection at two X-nulls on either side of the O-null. Figure \ref{fig:All_nulls} shows the radial locations of X- (black crosses) and O-nulls (red circles) as functions of time for a small angular wedge within $\pm 5 \degree$ of the solar equator and radial heights in the range $r = 1.1 - 1.5R_{s}$. This wedge encloses the lower part of the flare CS and the central part of the top of the flare arcade. 

Throughout the evolution, there is one main X-null (marked in Fig.\ \ref{fig:All_nulls}) in the flare CS from which most of the reconnection outflows, including both sunward- (negative slope in Fig.\ \ref{fig:All_nulls}) and outward-moving (positive slope in Fig.\ \ref{fig:All_nulls}) islands, are generated. The global evolution and the radially expanding geometry of the system dictate its location. Throughout the flare, the main null undergoes a modest change in height, much less than the change in height of the core O-null of the CME (also marked in Fig.\ \ref{fig:All_nulls}; it quickly moves out beyond the height range in the figure). Occasionally, an adjacent X-null becomes the new main X-null; one such transition occurs near $t=90,500$s in Figure \ref{fig:All_nulls}. According to \citet{Barta_2008_I, Barta_2011}, a newly formed island moves in the direction away from whichever of its delimiting X-nulls has the higher rate of reconnection, due to an excess of magnetic tension. If a second X-null forms at a radial distance above/below the main X-null, the island moves outward/sunward. \citet{Shen_2011} also report that the reconnection rate reaches a maximum at the main X-null in their MHD simulation of a flare CS. 

Sunward-moving islands merge with the flare arcade, which grows in height as the reconnected flux piles up. Therefore, these O-nulls stop their sunward motion at progressively higher altitudes; their last recorded positions trace approximately the instantaneous top of the arcade. The first outward-moving island forms the core of the flux rope that becomes the CME (CME core O-null in Fig.\ \ref{fig:All_nulls}). Subsequent outflowing islands merge with the rising flux rope above the maximum radial range of Figure \ref{fig:All_nulls}. We find that our islands exhibit little coalescence between them as they move along the flare CS in either direction. Instead, each island remains discrete until it merges with the flare arcade below the sheet or the CME flux rope above. 

\begin{figure}[h!]
   \centering
   \begin{center}$
      \begin{array}{cc}
          \resizebox{5.0in}{!}{\includegraphics{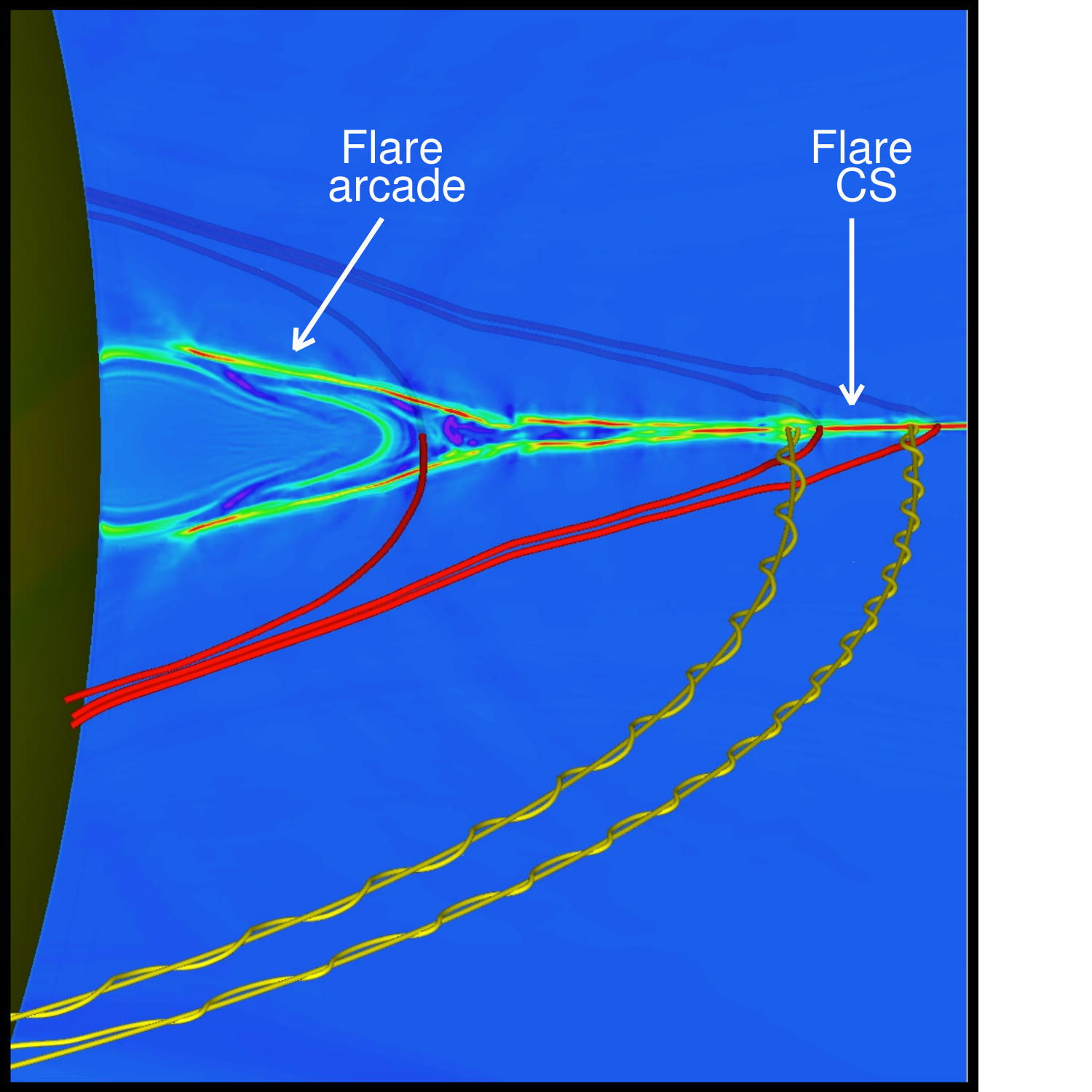}}
     \end{array}$
   \end{center}
\caption {Islands are disconnected flux ropes (yellow field lines) formed from two reconnection episodes at paired X-nulls. Over- and underlying sheared flux surfaces (red field lines) formed from a single reconnection episode at one X-null. The spherical surface to the left is at $1$ $R_{s}$. Color shading in the plane of the flare arcade ($r,\theta$ plane) shows $R_{s}$ $J_{\phi}/c$, saturated at $-18$ G (red) and $+4$ G (magenta). This simulation snapshot is at $t=84380$ s.} 
 \label{fig:Flux_rope} 
\end{figure}

Due to the magnetic shear in the erupting central arcade, each of our islands is a twisted 2.5D flux rope rather than a closed planar loop. Two examples (yellow lines) are shown in Figure \ref{fig:Flux_rope}. In our simulations, the flux ropes are not rooted at the Sun due to the special symmetry of the system; in 3D, they would be rooted on either side of the PIL, with finite displacements between their footpoints. As we noted above, the islands form due to reconnections of the same flux surfaces at two X-nulls, one above and one below the island. Beyond the ends of each island are flux surfaces that have reconnected only once at an X-null, which therefore remain rooted at the Sun. Three examples (red lines) are shown in Figure \ref{fig:Flux_rope}.

\subsection{Island Evolution}
\label{sec:island_evol}

\begin{figure}[h!]
  \centerline{\includegraphics{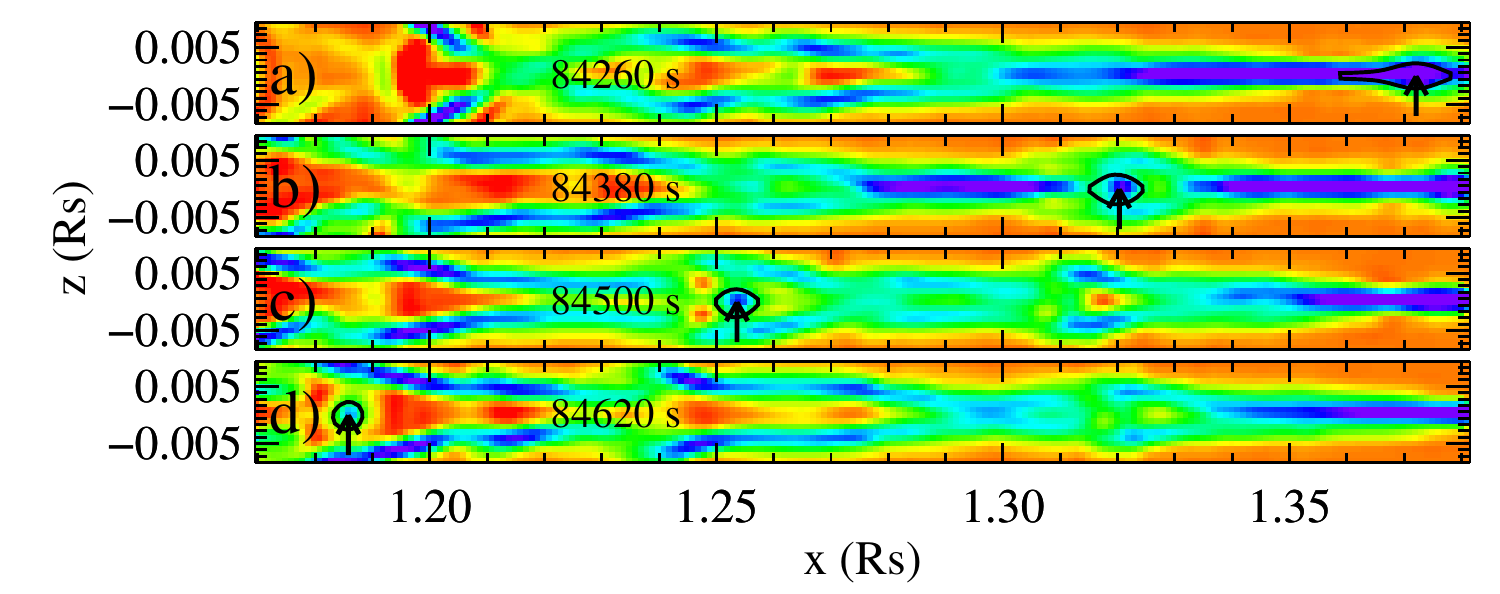}} 
   \caption {Normalized out-of-plane current density $R_{s}J_{\phi}/c$ at selected times for Island 1, in the $\phi = 0$ plane ($x = r \cos \theta, z = r \sin \theta$). Vertical black arrows point to the $x$ location of the island's O-null. A single flux surface is plotted as a closed black line in each frame. $R_{s}J_{\phi}/c$ is color-shaded from violet (saturated at $-17$ G) to red (saturated at $+2$ G). Images have been smoothed with a window of three pixels to minimize pixelation.}
 \label{fig:Image_sequence_1}
\end{figure}

\begin{figure}[h!]
   \centering
           \resizebox{4.0in}{!}{\includegraphics{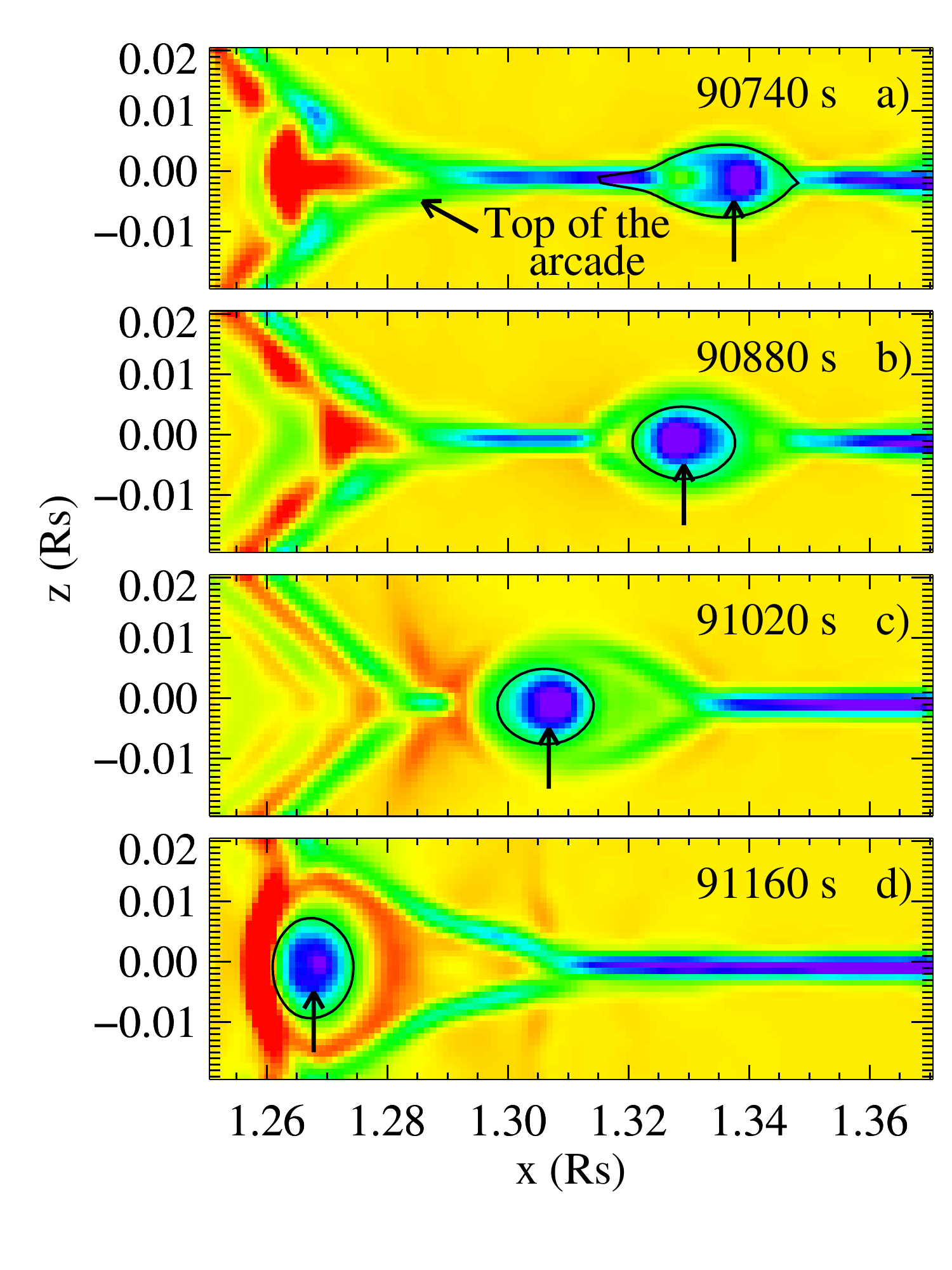}}
     \caption {Normalized out-of-plane current density $R_{s}J_{\phi}/c$ at selected times for Island 2, in the $\phi = 0$ plane ($x = r \cos \theta, z = r \sin \theta$). Vertical black arrows point to the $x$ location of the island's O-null. A single flux surface is plotted as a closed black line in each frame. $R_{s}J_{\phi}/c$ is color-shaded from violet (saturated at $-15$ G) to red (saturated at $+4$ G). Images have been smoothed with a window of three pixels to minimize pixelation.} 
 \label{fig:Image_sequence_2} 
\end{figure}

For this initial investigation of particle acceleration by island contraction following flare reconnection, we focus on the sunward-moving islands in the CS.  In particular, ``Island 1'' and ``Island 2'' in Figure \ref{fig:All_nulls} are tracked carefully from their birth to their arrival at the top of the flare arcade. We selected them because they are long-lived, clear features, with different physical characteristics. Several snapshots of these islands in the $\phi = 0$ plane ($x = r \sin \theta, z = r \cos \theta$) are shown in Figures \ref{fig:Image_sequence_1} and \ref{fig:Image_sequence_2}, with the x-locations of their respective O-nulls marked by black arrows. Island 1 is smaller and less resolved than Island 2. In Figure \ref{fig:Image_sequence_1}, a chain of islands can be seen both leading and trailing Island 1. The distortion of Island 2 during its interaction with the top of the arcade can be seen in Figure \ref{fig:Image_sequence_2}d. Each island has finite length along the CS and contains finite flux. As time passes and the reconnection continues, the flux entrained by the islands increases, so the islands appear to expand. The individual flux surfaces within each island contract with time, however, as demonstrated below.

The poloidal magnetic flux function, $\Psi(r,\theta)$, at each grid point, is
  \begin{equation}
  \label{eq:potential_vector}
        \Psi(r,\theta) = r \sin\theta A_\phi = r^2 {\int_0^\theta} d\theta'  \sin\theta' B_r(r,\theta'),
   \end{equation}
where $A_\phi(r,\theta)$ is the $\phi$ component of the magnetic vector potential. Isosurfaces of $\Psi$ trace the projection of field lines onto the computational $r-\theta$ plane because the magnetic field vector is tangential to those surfaces. Therefore, contours of $\Psi$ define individual flux surfaces. One sample flux surface (black line) is tracked versus time in each of Figures \ref{fig:Image_sequence_1} and \ref{fig:Image_sequence_2}.  

\begin{figure}[h!]
   \centering
   \begin{center}$
      \begin{array}{cc}
         \resizebox{6.0in}{!}{\includegraphics{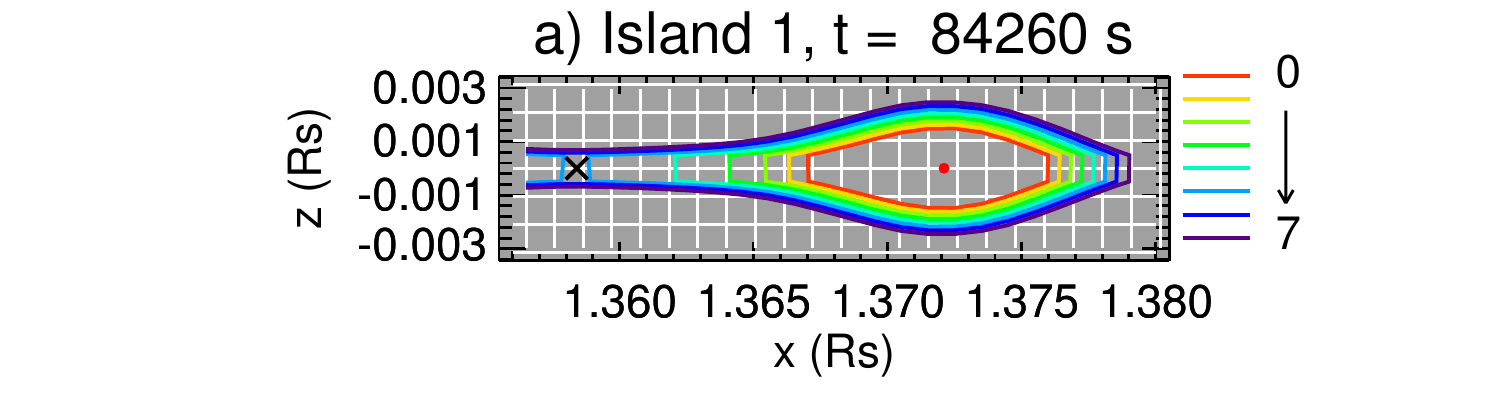}} \\
         \resizebox{6.0in}{!}{\includegraphics{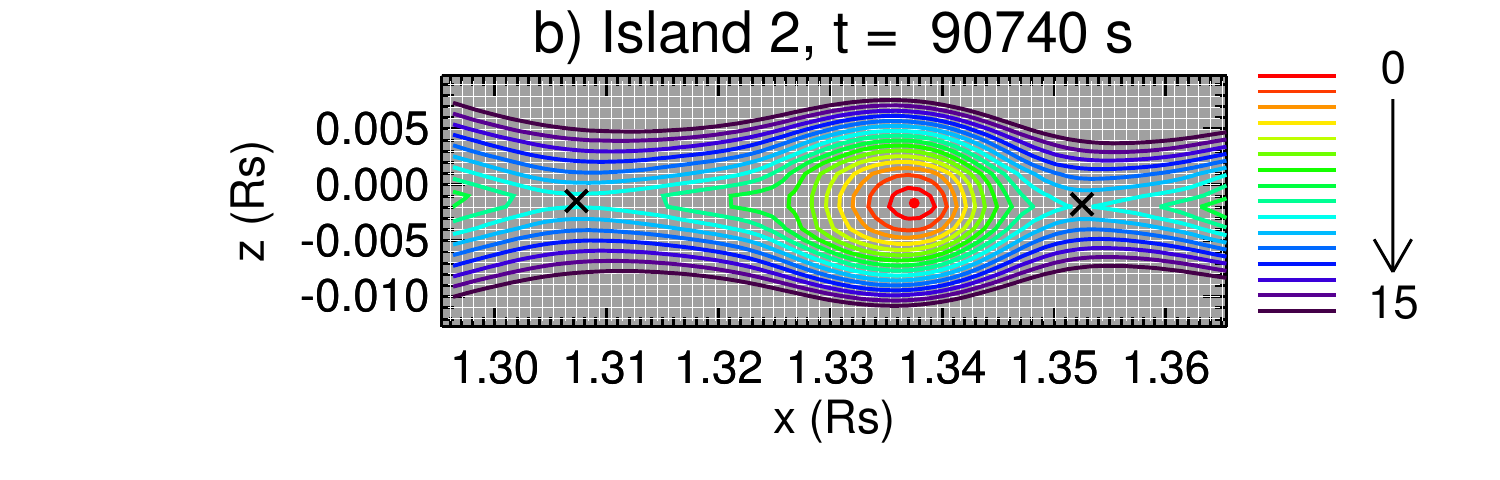}}
      \end{array}$
   \end{center}
\caption {Color-coded selected flux surfaces (curves) and null locations. Black crosses (red circles) are X-null (O-null) locations. The simulation grid is shown in white. a) Island 1 at $t=84260$s. Flux surfaces are colored red to black and numbered from $0$ to $7$, from innermost to outermost with respect to the island's O-null. The flux surface drawn in black in Figure \ref{fig:Image_sequence_1} is labeled ``5'' here. b) Island 2 at $t=90740$s. Flux surfaces are numbered from $0$ to $15$. The flux surface in Figure \ref{fig:Image_sequence_2} is labeled ``8'' here. Animations are available in the online journal (f7a\_movie.mpg and f7b\_movie.mpg).} 
 \label{fig:Flux_surfaces} 
\end{figure}

To determine the characteristic properties of our islands over time, we chose a set of fixed flux surfaces with selected values of $\Psi$ inside each island. We tracked each surface from its creation at time $t_{i}$ to its arrival at the flare arcade. Figure \ref{fig:Flux_surfaces} shows the set of chosen flux surfaces (color-coded and numbered) at a time well after their formation for Island 1 (\ref{fig:Flux_surfaces}a, top panel) and Island 2 (\ref{fig:Flux_surfaces}b, bottom panel). The instantaneous locations of the island nulls are marked by black crosses (X-nulls) and red circles (O-nulls). The X-null to the right of Island 1 is not shown because it is well outside the range of the plot, at $x=1.3984$. The number of flux surfaces chosen for Island 1 is smaller than for Island 2 because of the difference in size. The grid lines (white) show that the innermost flux surface ($0$) of Island 1 is only about three cells across, the minimum resolvable area, and the enclosed areas of flux surfaces $0$ and $1$ are not resolved after time $t \approx 84440$ due to contraction. Although our simulation uses the highest resolution to date, resolving the islands, particularly their innermost flux surfaces that form earliest, is a significant computational challenge.

The flux surfaces within each island are created sequentially, from innermost (closest to the O-null) to outermost, so $t_{i}$ generally increases across our sample of flux surfaces. These initial times are listed in Tables \ref{table:Island_1} and \ref{table:Island_2} for Islands 1 and 2, respectively. The cadence of output from the simulation run was $20$ s, which is the uncertainty in $t_{i}$.

\begin{table}[ht]
\begin{tabular}{c c c c c c c c c c c}                   
\hline \hline                                       

\shortstack{ $\Psi$\# } & \shortstack{ $t_{i}$ \\ ($10^{4}$ s) } & \shortstack{ $L_{p,i}$ \\ $(R_{s})$ }  & \shortstack{ $L_{i}$ \\ $(R_{s})$ } & \shortstack{ $B_{1i}$ \\ (G) } & 
\shortstack{ $B_{2i}$ \\ (G) } &  \shortstack{$\left<n_{i}\right>$ \\ ($10^{7}$ cm$^{-3}$)} & \shortstack{ $\left<B_{p,i}^{2}\right>$ \\ (G$^{2}$) } & \shortstack{ $\left<B_{t,i}^{2}\right>$ \\ (G$^{2}$) } & \shortstack{ $\left< 8 \pi P_{i} \right>$ \\ (G$^{2}$) } &  \shortstack{ $\mathcal{E}_{max}$ } \\  [0.5ex]
\hline
 0  &  8.422  &   0.036  &   0.125  &   0.423  &   0.661  &   0.712  &   0.049  &   0.265  &   0.043  &   3.817  \\
 1  &  8.424  &   0.030  &   0.099  &   0.492  &   0.718  &   0.849  &   0.058  &   0.334  &   0.057  &   2.183  \\
 2  &  8.424  &   0.034  &   0.106  &   0.475  &   0.715  &   0.812  &   0.064  &   0.319  &   0.053  &   2.260  \\
 3  &  8.424  &   0.038  &   0.114  &   0.451  &   0.710  &   0.782  &   0.069  &   0.303  &   0.050  &   2.375  \\
 4  &  8.426  &   0.034  &   0.114  &   0.494  &   0.730  &   0.873  &   0.070  &   0.349  &   0.060  &   2.153  \\
 5  &  8.426  &   0.041  &   0.139  &   0.457  &   0.727  &   0.812  &   0.070  &   0.322  &   0.053  &   2.812  \\
 6  &  8.430  &   0.034  &   0.127  &   0.561  &   0.760  &   0.990  &   0.074  &   0.397  &   0.071  &   1.874  \\
 7  &  8.432  &   0.046  &   0.217  &   0.539  &   0.764  &   1.025  &   0.053  &   0.390  &   0.074  &   4.718  \\
 [1ex] 
 \hline
\end{tabular}
\caption{Island 1. First column: Flux surfaces as labeled in Figure \ref{fig:Flux_surfaces}a. ``$< >$'' indicates quantities averaged along the perimeter of the flux surface.}
\label{table:Island_1}
\end{table}

\begin{figure}[h!]
   \centering
   \begin{center}$
      \begin{array}{cc}
           \resizebox{3.0in}{!}{\includegraphics{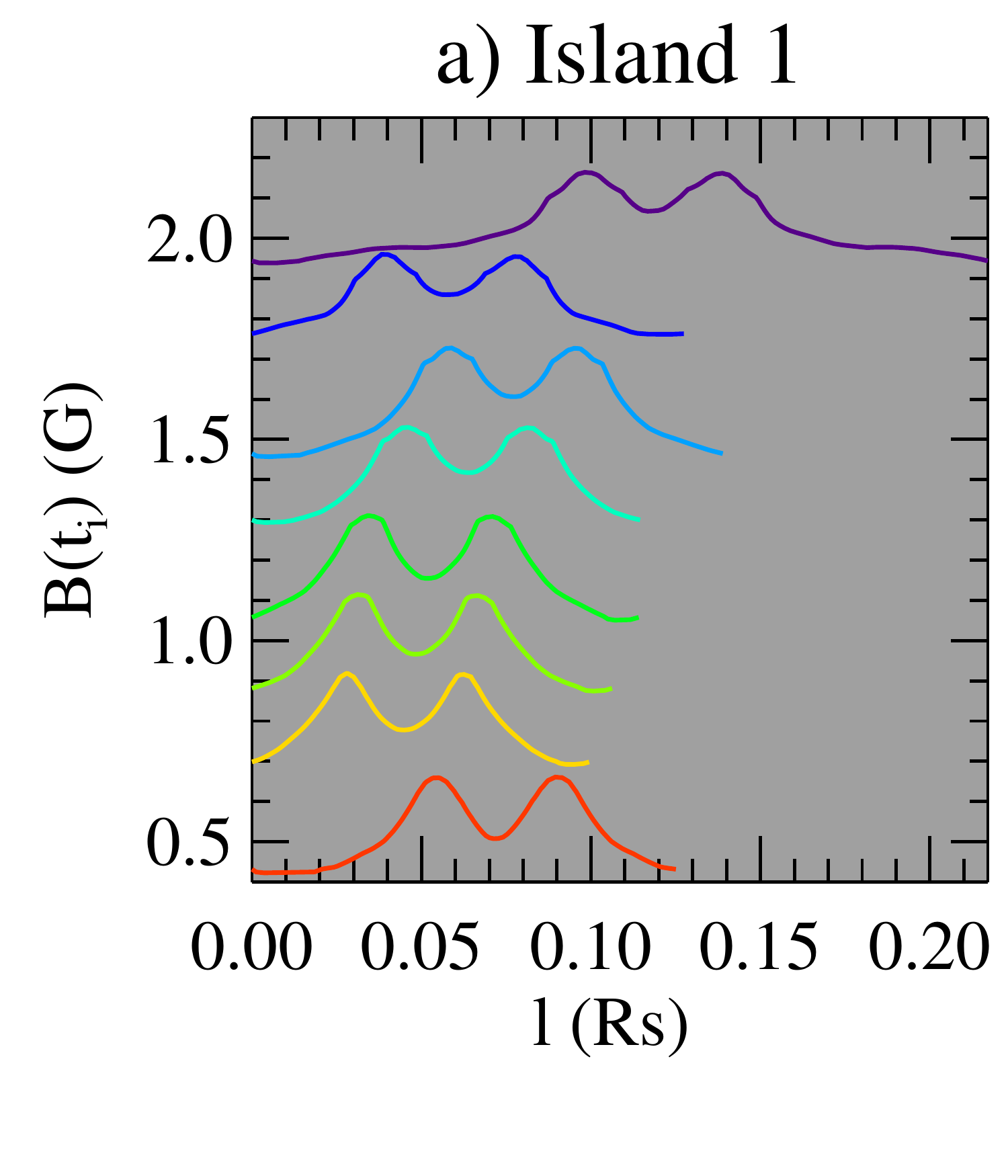}}
            \resizebox{3.0in}{!}{\includegraphics{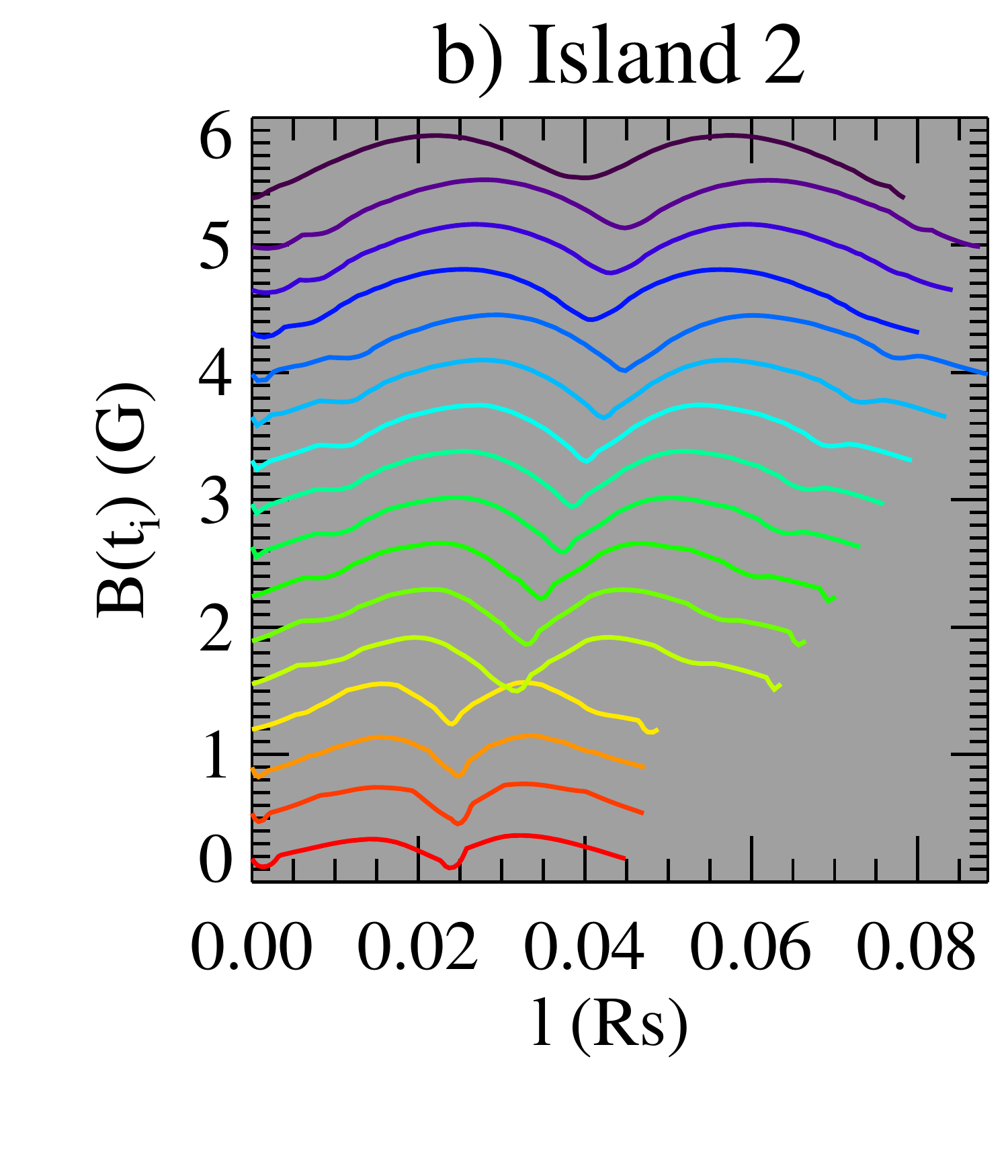}}
      \end{array}$
   \end{center}
\caption {Magnetic field magnitude versus arc length of the flux surfaces color-coded in Figure \ref{fig:Flux_surfaces} at the formation time $t_{i}$ listed in Tables \ref{table:Island_1} and \ref{table:Island_2}. a): Island 1; b) Island 2. To separate the profiles, each curve has been shifted upward by $0.20$ G (Island 1) or $0.35$ G (Island 2) from the curve below.} 
 \label{fig:B_arc_length} 
\end{figure}

The profile of magnetic field strength versus arc length along each flux surface at time $t_{i}$ is shown in Figure \ref{fig:B_arc_length}.  Arc length $l=0$ corresponds to the point of the field line closest to the Sun's surface, and the end of each curve corresponds to one turn of the field line at $l=L$. The minima of $B$ ($B_{1}$) occur near the two X-nulls, and the two maxima ($B_{2}$) occur between those points, consistent with the analytical formula we adopted in Equation \ref{eq:B_prof_res}. The inhomogeneity is more pronounced for Island 2, whose poloidal (in-plane) field component $B_{p}$, is strongly dominant: the toroidal (out-of-plane) field component $B_{t}$ weakens with time as less-sheared flux reconnects across the flare CS, so the island's bounding X-nulls are closer to true nulls of the magnetic field. Tables \ref{table:Island_1} and \ref{table:Island_2} list the average along the perimeter of each flux surface of the squared toroidal and poloidal field strengths. The sinusoidal profile flattens with time for each flux surface, until the island collides with the arcade.

\begin{table}[h!]
\begin{tabular}{c c c c c c c c c c c}                   
\hline \hline                                       

\shortstack{ $\Psi$\# } & \shortstack{ $t_{i}$ \\ ($10^{4}$ s) } & \shortstack{ $L_{p,i}$ \\ $(R_{s})$ }  & \shortstack{ $L_{i}$ \\ $(R_{s})$ } & \shortstack{ $B_{1i}$ \\ (G) } & 
\shortstack{ $B_{2i}$ \\ (G) } &  \shortstack{$\left<n_{i}\right>$ \\ ($10^{7}$ cm$^{-3}$)} & \shortstack{ $\left<B_{p,i}^{2}\right>$ \\ (G$^{2}$) } & \shortstack{ $\left<B_{t,i}^{2}\right>$ \\ (G$^{2}$) } & \shortstack{ $\left< 8 \pi P_{i} \right>$ \\ (G$^{2}$) } &  \shortstack{ $\mathcal{E}_{max}$ } \\  [0.5ex]
\hline
 0  &  9.060  &   0.039  &   0.045  &   0.111  &   0.365  &   1.173  &   0.069  &   0.013  &   0.124  &   5.186  \\
 1  &  9.062  &   0.042  &   0.047  &   0.106  &   0.420  &   1.178  &   0.093  &   0.014  &   0.127  &   3.687  \\
 2  &  9.064  &   0.043  &   0.047  &   0.124  &   0.449  &   1.170  &   0.103  &   0.013  &   0.127  &   3.316  \\
 3  &  9.066  &   0.044  &   0.049  &   0.125  &   0.516  &   1.156  &   0.130  &   0.014  &   0.124  &   3.376  \\
 4  &  9.066  &   0.059  &   0.064  &   0.099  &   0.520  &   0.947  &   0.133  &   0.009  &   0.089  &   4.318  \\
 5  &  9.068  &   0.063  &   0.067  &   0.111  &   0.546  &   0.924  &   0.148  &   0.009  &   0.085  &   3.859  \\
 6  &  9.070  &   0.066  &   0.070  &   0.105  &   0.560  &   0.905  &   0.156  &   0.009  &   0.082  &   4.076  \\
 7  &  9.072  &   0.070  &   0.073  &   0.102  &   0.567  &   0.896  &   0.162  &   0.008  &   0.081  &   4.169  \\
 8  &  9.074  &   0.072  &   0.076  &   0.094  &   0.581  &   0.892  &   0.168  &   0.008  &   0.080  &   4.444  \\
 9  &  9.076  &   0.076  &   0.079  &   0.089  &   0.594  &   0.881  &   0.174  &   0.008  &   0.078  &   4.662  \\
10  &  9.078  &   0.080  &   0.083  &   0.084  &   0.597  &   0.871  &   0.176  &   0.008  &   0.077  &   4.903  \\
11  &  9.080  &   0.085  &   0.088  &   0.086  &   0.601  &   0.858  &   0.177  &   0.007  &   0.074  &   4.600  \\
12  &  9.084  &   0.076  &   0.080  &   0.077  &   0.609  &   0.961  &   0.193  &   0.009  &   0.090  &   4.840  \\
13  &  9.086  &   0.079  &   0.084  &   0.075  &   0.612  &   0.960  &   0.193  &   0.009  &   0.090  &   4.568  \\
14  &  9.088  &   0.082  &   0.087  &   0.073  &   0.612  &   0.958  &   0.193  &   0.009  &   0.089  &   4.136  \\
15  &  9.092  &   0.075  &   0.078  &   0.118  &   0.609  &   1.077  &   0.207  &   0.011  &   0.108  &   2.317  \\
 [1ex] 
\hline
\end{tabular}
\caption{Island 2. First column: flux surfaces as labeled in Figure \ref{fig:Flux_surfaces}b. ``$< >$'' indicates quantities averaged along the perimeter of the flux surface.}
\label{table:Island_2}
\end{table}

\begin{figure}[h!]
 \centering
   \resizebox{6.0in}{!}{\includegraphics{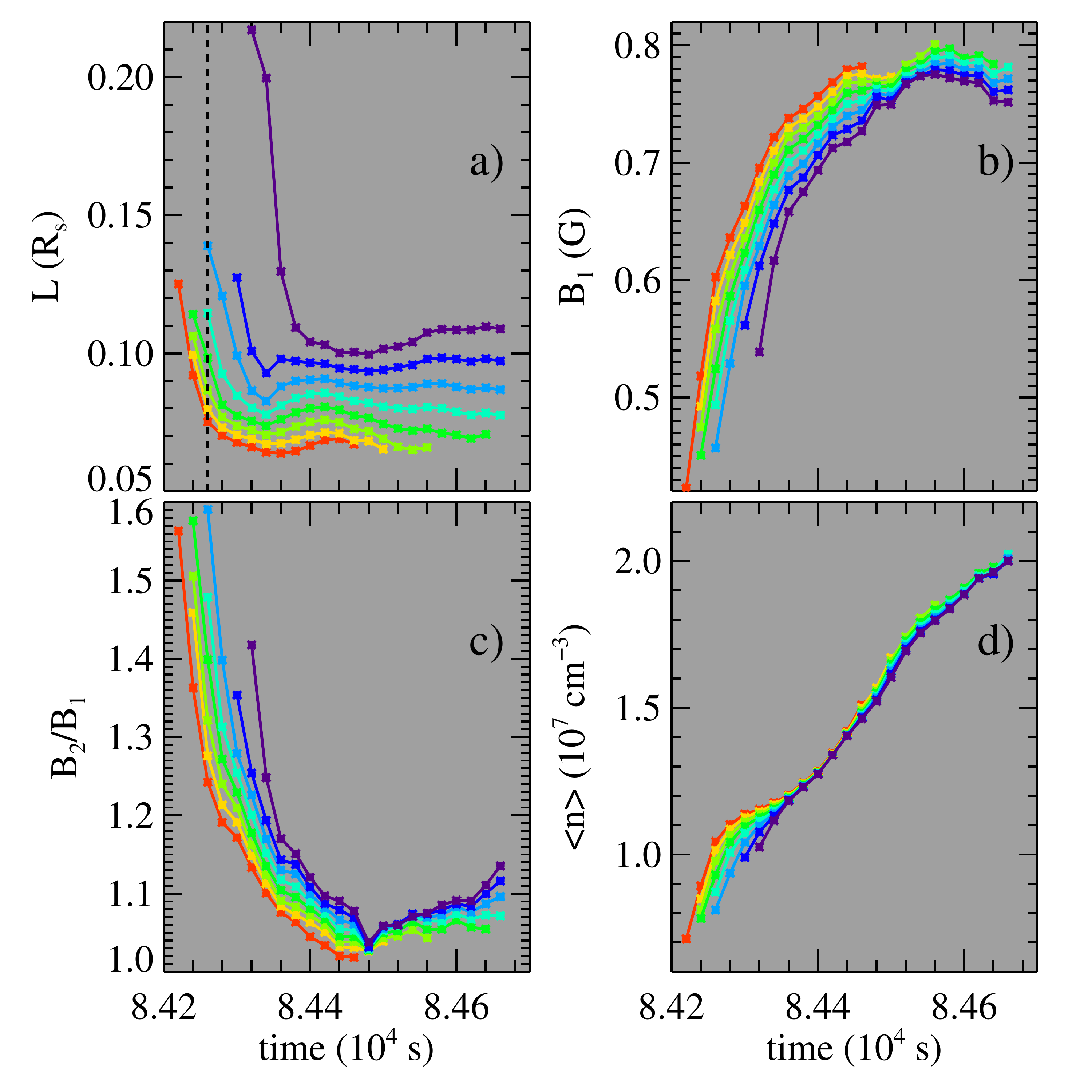}}
    \caption {Island 1. Evolution of properties of the flux surfaces in Figure \ref{fig:Flux_surfaces}a, using the same color code; the vertical dashed line marks the time shown in that figure. a) Length of one turn of a field line, $L$. b) Minimum magnetic field strength, $B_{1}$. c) Mirror ratio, $B_{2}/B_{1}$. d) Electron density averaged along the perimeter of the flux surface, $\left< n \right>$.}
  \label{fig:Island1_macroscopic}  
\end{figure}

\begin{figure}[h!]
 \centering
   \resizebox{6.0in}{!}{\includegraphics{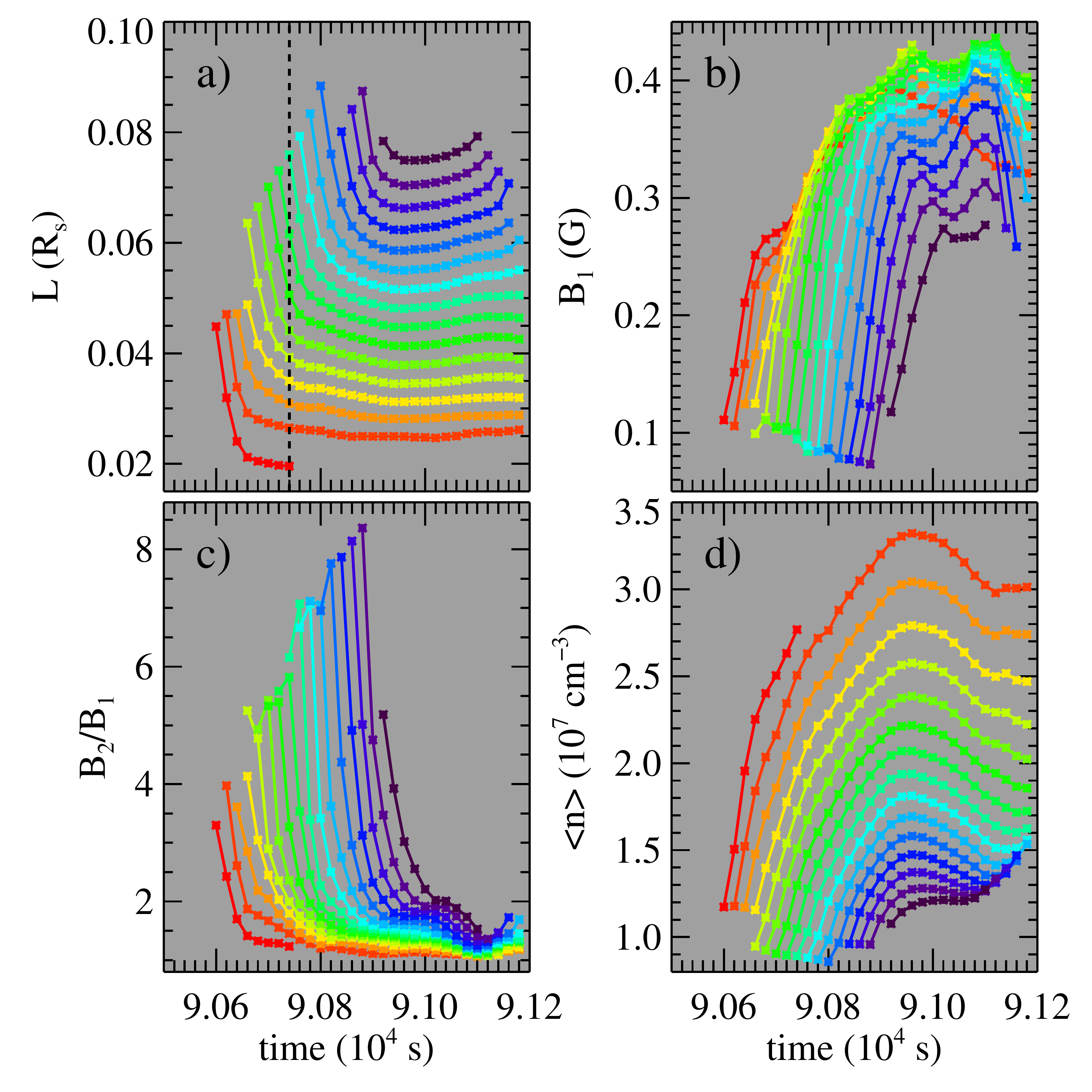}}
    \caption {Island 2. Evolution of properties of the flux surfaces in Figure \ref{fig:Flux_surfaces}b, using the same color code; the vertical dashed line marks the time shown in that figure. a) Length of one turn of a field line, $L$. b) Minimum magnetic field strength, $B_{1}$. c) Mirror ratio, $B_{2}/B_{1}$. d) Electron density averaged along the perimeter of the flux surface, $\left< n \right>$.}
  \label{fig:Island2_macroscopic}  
\end{figure}

For each flux surface, we measured the properties required to determine a particle's speed versus time (\S \ref{sec:analytical_A}): $L$, $B_{1}$, and $B_{2}$. $L$ is calculated from the arc-length integral along the perimeter of each flux surface, 
\begin{equation}
      \label{eq:arc_l}
       L = \int dl =  \int \left( 1 + \frac{B^2_{t}}{B^2_{p}} \right)^{1/2} dl_{p},
 \end{equation}
where $dl_{p}$ is the 2D arc-length differential in the poloidal $r-\theta$ plane. $L_{p} = \int dl_{p}$ is the perimeter of a flux surface, while $L $ includes the extra length of one turn of the associated field line due to its out-of-plane component of the field. The variations of $L$ over time are displayed in Figures \ref{fig:Island1_macroscopic}a (for Island 1) and \ref{fig:Island2_macroscopic}a (for Island 2). Values are not plotted for those times when flux surfaces are unresolved, i.e., when $L_{p}$ is smaller than the perimeter of a $2 \times 3$-cell box. All flux surfaces experience a sharp decrease in length soon after formation, when they are elongated along the CS direction. Thereafter, the changes in lengths are much smaller.

For Island 1, the maximum of flux surfaces ratios $L/L_{i}$ ranges between $1.4$ and $2.0$, with the exception of the outermost flux surface (flux surface $7$, violet color), where of $L/L_{i} =2.19$. Flux surface $7$ is initially much longer than its neighbors because at $t_{i}$ it encloses an extra X-null from a small, short-lived island that coalesces with Island 1 between $84300$ and $84360$ s. This flux surface initially has a large aspect ratio and then retracts, reducing its length quickly. This shows the effect of island coalescence in enhancing island contraction. Flux surfaces that were inside Island 1 before the merger experience a temporary increase in their length as they become part of the reconnection upstream flow along the direction of the CS, at the null between the islands. There is another coalescence with a much smaller and short-lived island between $84400$ and $84440$ s, with insignificant effects on the chosen flux surfaces.

Island 2 has similar behavior to Island 1, except that it does not merge with other islands; it only interacts with the flare arcade. The maximum ratio $L/L_{i}$ for Island 2 is between $\approx 1.05$ (outermost flux surface) and $2.26$ (innermost flux surface). All flux surfaces are well-resolved at all times except flux surface ``0'', whose enclosed area is not resolved after time $t \approx 90740$ (see Fig.\ \ref{fig:Flux_surfaces}b for the shape of this flux surface at this time). At $t \approx 90900$s, the X-null closest to the Sun starts to move quickly toward the Sun leaving the opposite X-null behind, which in turn becomes the new main X-null of the flare (see Fig.\ \ref{fig:All_nulls}). This separation of the nulls slightly stretches the flux surfaces.
 
Island 1 has a much smaller poloidal area than Island 2, but the lengths of its field lines are in general much larger because its toroidal field is so strong relative to its poloidal field. This can be seen, for example, in the listed values of $L_{p}$ and $L$ at time $t_{i}$ in Tables \ref{table:Island_1} and \ref{table:Island_2}. The toroidal component of the field increases almost monotonically with time for the flux surfaces of Island 1 as the total magnetic pressure increases. For Island 2, this component increases until the sunward X-null starts moving rapidly toward the flare arcade, then it decreases as the total thermal pressure decreases. Tables \ref{table:Island_1} and \ref{table:Island_2} list the initial magnetic (separated by components) and thermal pressures of each flux surface. Island 1 is magnetically dominated, and its plasma-$\beta$ decreases with time. In contrast, Island 2 is thermally dominated, and its plasma-$\beta$ increases with time.
 
A particle's parallel and perpendicular energy gains directly depend on the evolution of $B_{1}$ (Eqs. \ref{eq:E_perp_results} and  \ref{eq:E_paral_results}). The evolution of $B_{1}$ is shown in Figs. \ref{fig:Island1_macroscopic}b (Island 1) and \ref{fig:Island2_macroscopic}b (Island 2). For both islands, $B_{1}$ increases rapidly at first as the islands tend to a more uniform magnetic field, then much more slowly, with occasional fluctuations. In general, the evolution of $B_{1}$ is favorable for both components of energy gain. The homogenization of the field can be seen in Figs. \ref{fig:Island1_macroscopic}c and \ref{fig:Island2_macroscopic}c, where the mirror ratio of flux surfaces is plotted.  For each island, $B_{2}/B_{1}$ decreases rapidly as $B_{1}$ increases; later, the rate of decrease lessens until the mirror ratio fluctuates mildly. For Island 1, its flux surfaces are most uniform at about $84480$ s. After this time, the magnetic field profile no longer resembles the analytical form proposed in \S \ref{sec:analytical_A} (Eq.\ \ref{eq:B_prof_res}); energy gains are not calculated after this time in \S \ref{sec:part_en_gain}. Island 2 exhibits similar behavior: the field is most uniform at about $91080$ s, after which our assumptions on the shape of the magnetic field profile are no longer applicable. Island 2's $B_{1}$ temporarily decreases after its sunward X-null moves quickly toward the Sun.

The peak mirror ratio in Island 1 is $\approx 1.6$, much smaller than in Island 2, which achieves a value as large as $\approx 8$. The difference in the maximum mirror ratios is due to the moderating effect of the toroidal field and the aspect ratio of the flux surfaces. Island 1 has a strong toroidal field and behaves closer to the uniform-field case discussed in Appendix \ref{sec:appen_B}. In contrast, Island 2 has a weak toroidal field, so it behaves like a strong magnetic trap initially and transitions to uniform-field behavior very late, relative to Island 1.  For Island 2, the outermost flux surfaces have the largest mirror ratios because the field strength minima always are very small near the X-nulls, whereas the maxima are highest at the greatest cross-sheet distance from the O-null. Consequently, the mirror ratio for newly formed flux surfaces more than doubles over time for Island 2 (Fig.\ \ref{fig:Island2_macroscopic}c), whereas it declines slightly, but irregularly, for Island 1 (Fig.\ \ref{fig:Island1_macroscopic}c). The initial values of the minimum and maximum field strengths on each flux surface are listed in Tables \ref{table:Island_1} and \ref{table:Island_2}. 

Plasma compression occurs as the islands contract. In the uniform-field limit discussed in Appendix \ref{sec:appen_B}, we show that plasma compression contributes directly to the perpendicular energy gain. Figures \ref{fig:Island1_macroscopic}d and \ref{fig:Island2_macroscopic}d show the electron density averaged over the flux surfaces, $<n>$, as functions of time. For all flux surfaces in Island 1, the electron density increases monotonically and has similar values for all flux surfaces (due mainly to their proximity in this small island). The electron density in Island 2 increases monotonically until $t \approx 91000$ s (when expansion of the island starts) and then decreases. The density ratio for both islands $\left< n \right> / \left< n_{i} \right>$ ranges from $1$ to $\approx 2.8$, with the innermost surfaces having the highest compression ratios. The initial densities, $\left< n_{i} \right>$, are listed in Tables \ref{table:Island_1} and \ref{table:Island_2}.

We estimated the number of electrons per radian of angular extent out of the plane, $N'$, enclosed by the outermost studied flux surface of each island. We integrated the electron number density across the area enclosed by that flux surface, and then multiplied this integral by $R_{s}$ to account for the out-of-plane length. For the smaller Island 1, $N' \approx 2\times10^{35}$ electrons per radian; for the larger Island 2, $N' \approx 2.5 \times10^{36}$ electrons per radian. Thus, our results suggest that it is theoretically possible to accelerate large numbers of electrons in flares by this macroscopic process. 

\subsection{Particle Energy Gain}
\label{sec:part_en_gain}

\begin{figure}[h!]
 \centering
   \resizebox{6.0in}{!}{\includegraphics{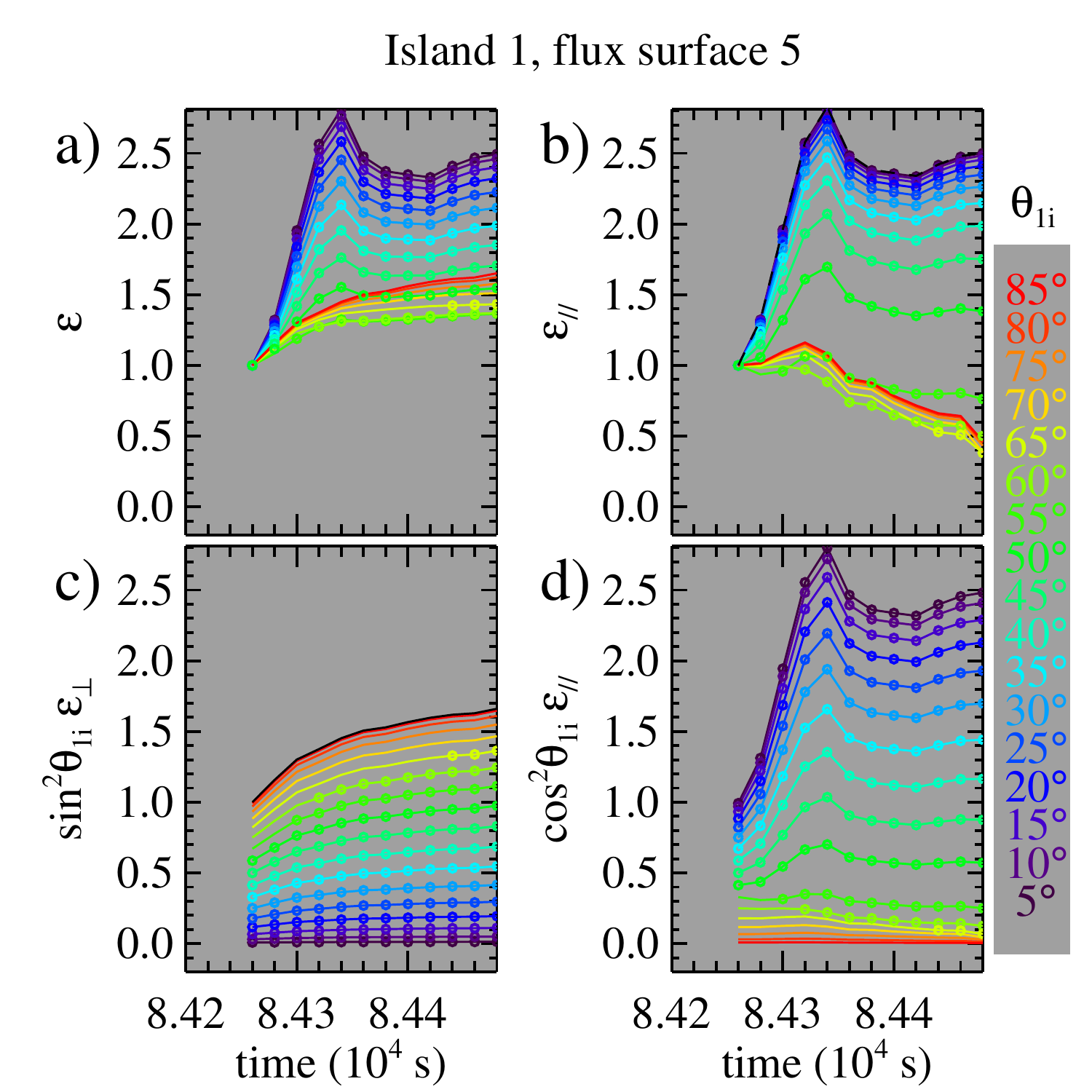}}
    \caption {Island 1, flux surface $5$. Energy ratios, color-coded by initial pitch angles [$5^\circ$,$85^\circ$] listed at right. a) Total energy ratio, $\mathcal{E}$, from Equation (\ref{eq:en_gain_results}). b) Parallel energy ratios, $\mathcal{E}_{\parallel}$, from Eq. (\ref{eq:E_paral_results}). The special case $\mathcal{E}_{\parallel} (\theta = 0) = (L_{i}/L)^2$ is shown in black (it closely follows the $\theta = 5^\circ$ case.) c) Contribution of perpendicular energy ratio to total, first term of Equation (\ref{eq:en_gain_results}). Perpendicular energy ratio, $\mathcal{E}_{\perp}$,  from Eq.(\ref{eq:E_perp_results}), is shown with a solid black curve.  d) Contribution of parallel energy ratio to total, $\mathcal{E}_{\parallel}$ to $\mathcal{E}$, second term of Equation (\ref{eq:en_gain_results}). Except for the solid black lines, points overplotted with a circle indicate transiting motion, and those without a circle indicate mirroring motion. An animation showing the energy ratios for all flux surfaces of Island 1 is included in the online journal (f11\_movie.mpg).}  
\label{fig:Island1_energy_ratios}
\end{figure}

\begin{figure}[h!]
 \centering
   \resizebox{6.0in}{!}{\includegraphics{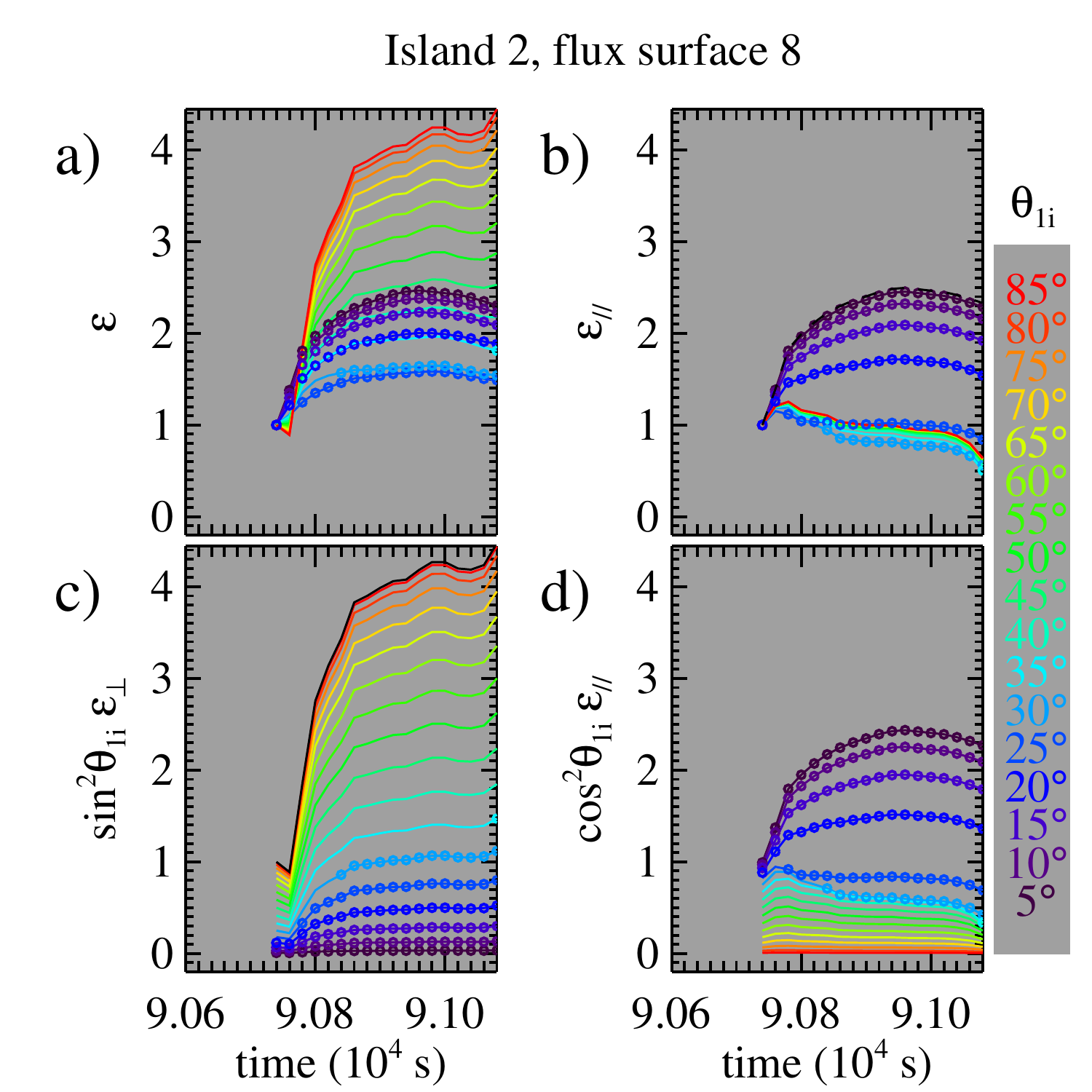}}
    \caption {Island 2, flux surface $8$. Energy ratios, color-coded by initial pitch angles [$5^\circ$,$85^\circ$] listed at right. a) Total energy ratio, $\mathcal{E}$, from Equation (\ref{eq:en_gain_results}). b) Parallel energy ratios, $\mathcal{E}_{\parallel}$, from Eq. (\ref{eq:E_paral_results}). The special case $\mathcal{E}_{\parallel} (\theta = 0) = (L_{i}/L)^2$ is shown in black (it closely follows the $\theta = 5^\circ$ case.) c) Contribution of perpendicular energy ratio to total, first term of Equation (\ref{eq:en_gain_results}). Perpendicular energy ratio,  $\mathcal{E}_{\perp}$,  from Eq.(\ref{eq:E_perp_results}), is shown with a solid black curve.  d) Contribution of parallel energy ratio to total, $\mathcal{E}_{\parallel}$ to $\mathcal{E}$, second term of Equation (\ref{eq:en_gain_results}). Except for the solid black lines, points overplotted with a circle indicate transiting motion, and those without a circle indicate mirroring motion. An animation showing the energy ratios for all flux surfaces of Island 2 is included in the online journal (f12\_movie.mpg).}  
\label{fig:Island2_energy_ratios}
\end{figure}

We now calculate the energy gain of particles within our islands using the properties from \S \ref{sec:island_evol} and Equations (\ref{eq:en_gain_results})--(\ref{eq:E_paral_results}). As discussed previously, the energy gain of a particle moving adiabatically along a flux surface is a function of the particle's initial pitch angle and the properties of the flux surface; it is independent of the particle's initial energy.  Therefore, for each flux surface in Islands 1 and 2, we calculated the energy gain for particles with initial pitch angles $\theta_{1i}$ in the range $5\degree$ to $85\degree$, at $5\degree$ intervals. At selected times after formation time $t_{i}$, we used the instantaneous properties of the flux surfaces described in \S \ref{sec:island_evol}, solved the transcendental equations derived in Appendix \ref{sec:appen_A}, and calculated the new pitch angles and the energy ratios. The value of $\theta$ and the mirror ratio of the flux surface a particle is orbiting determine whether the particle is transiting or mirroring. Conversions between these two populations occur as the flux surface properties change over time, as shown in the figures below and discussed in Appendix \ref{sec:appen_A}.
 
Results for a representative flux surface within Island 1 (black curves in Fig.\ \ref{fig:Image_sequence_1} and flux surface labeled ``5'' in Fig.\ \ref{fig:Flux_surfaces}a) are shown in Figure \ref{fig:Island1_energy_ratios}, and those for a flux surface within Island 2 (black curves in Fig.\ \ref{fig:Image_sequence_2} and flux surface labeled ``8'' in Fig.\ \ref{fig:Flux_surfaces}b) are shown in Figure \ref{fig:Island2_energy_ratios}. The colors represent the initial pitch angles, tabulated at the right of each figure. They range from violet at 5$^\circ$ for particles with almost solely parallel motion, which execute transiting orbits, to red at 85$^\circ$ for particles with almost solely perpendicular motion, which execute mirroring orbits. Points overplotted with a circle indicate transiting particles; those without a circle indicate mirroring particles. In both figures, particles at certain initial pitch angles (e.g., 65$^\circ$, light green, in Fig.\ \ref{fig:Island1_energy_ratios}; 35$^\circ$, medium blue, in Fig.\ \ref{fig:Island2_energy_ratios}) convert from mirroring to transiting motion as the mirror ratio of field strengths along the flux surface decreases over time. 

The principal results of this study are illustrated by Figs. \ref{fig:Island1_energy_ratios}a and \ref{fig:Island2_energy_ratios}a, which show the total energy ratios, $\mathcal{E}$, from Equation (\ref{eq:en_gain_results}). We find that the maximum total energy ratio is $\mathcal{E}_{max} \approx 4.7$ for Island 1, on the outermost flux surface, which is involved in the island coalescence (for the effects of energy gain by island coalescence in different regimes see, for example, \citet{Fermo_2010} and \citet{Zank_2014}). The last column of Table \ref{table:Island_1} lists $\mathcal{E}_{max}$ for each flux surface. In general, the values of $\mathcal{E}_{max}$ are larger for Island 2 (see Table \ref{table:Island_2}), even though this island does not experience any coalescence with other islands, achieving a maximum of $\approx 5.2$. These values are somewhat larger than the factor of $\approx 3$ reported by \citet{Drake_2010} for electrons orbiting a single contracting island in particle-in-cell (PIC) simulations of multiple, parallel CSs. Our values are greater due mainly to compression of the flux surfaces as our islands stream down the flare CS toward the top of the flare arcade. Most of Drake's islands in PIC simulations are incompressible, i.e. the area of islands is conserved after island merging (see \citealt{Fermo_2010}).

To understand the energy gains in greater detail, we display the parallel energy ratio, $\mathcal{E}_{\parallel}$ in Eq. (\ref{eq:E_paral_results}), in panel b) of Figures \ref{fig:Island1_energy_ratios} and \ref{fig:Island2_energy_ratios}. $\mathcal{E}_{\perp}$ (Eq. (\ref{eq:E_perp_results})) is shown with a black curve in panel c) of the same figures. The increasing strength of the magnetic field in the contracting islands increases the perpendicular energy of all particles, in order to conserve their magnetic moment. A steep rise in perpendicular energy gain occurs early, when $B_{1}$ changes rapidly, then the growth levels off while the islands contract slowly. The parallel energy of a particle will increase (decrease) if the length of its orbit decreases (increases), in order to conserve parallel action. For transiting particles, their parallel energy gain increases as long as the island contracts, independent of their pitch angle, because the length of their orbits is the length of the turn of the field line ($L$). The maximum parallel energy gain occurs for particles with $\theta = 0$, with $\mathcal{E}_{\parallel max} = (L_{i}/L)^2 $. Depending on how much $B_{1}$ changes compared to the change in the square of the length, $\mathcal{E}_{\perp}$ may be smaller (Island 1 case) or larger (Island 2 case) than $\mathcal{E}_{\parallel max}$. $L_{i}/L$ is similar for both islands, but the change in $B_{1}$ is larger for Island 2.

\begin{figure}[h!]
 \centering
   \resizebox{6.0in}{!}{\includegraphics{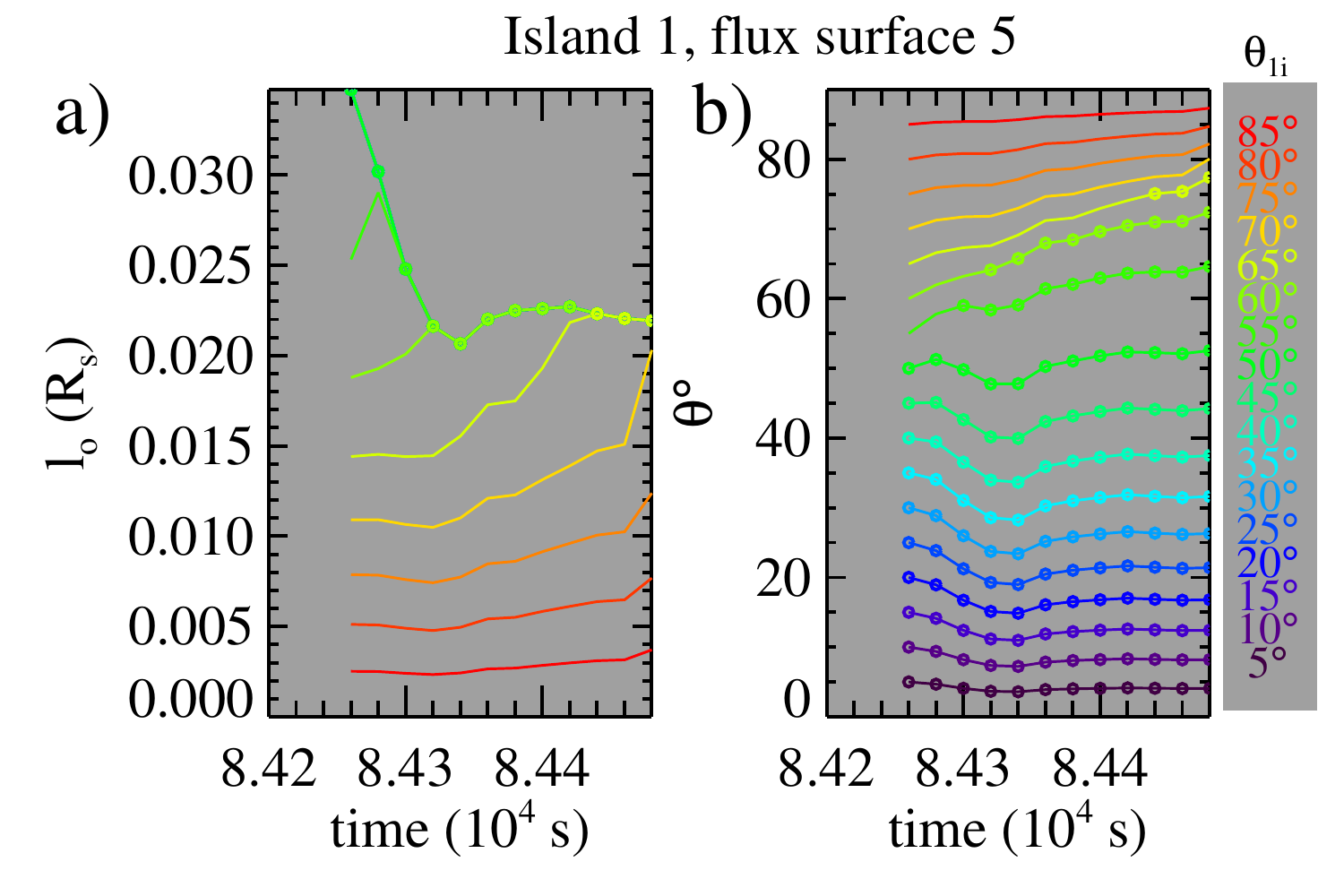}}
    \caption {Island 1, flux surface $5$. Evolution of $l_{o}$ and $\theta$ for the same color-coded initial pitch angles as in Figure \ref{fig:Island1_energy_ratios}. a) $l_{o}$. b) $\theta$. An animation showing $l_{o}$ and $\theta$ for all flux surfaces of Island 1 is included in the online journal (f13\_movie.mpg).}  
\label{fig:Island1_orbit_length}
\end{figure}

For mirroring particles, the length of their orbits may decrease (increase) if the separation of the ``mirrors'' decreases (increases), even if the total length of the flux surface decreases. This depends on how fast the mirror ratio, the pitch angle, and the total length of the field line change (see Eqs.\  \ref{eq:beta} and \ref{eq:orbit_length}). For example, Figs.\  \ref{fig:Island1_orbit_length} and \ref{fig:Island2_orbit_length} show $l_{o}$ (a quarter of the orbit length, see Appendix \ref{sec:appen_A} for details) for all particles of flux surfaces in Figs.\ \ref{fig:Island1_energy_ratios} and \ref{fig:Island2_energy_ratios}, respectively. For both islands, most of the mirroring particles experience a slight decrease in $l_{o}$ at early times, which translates into an small increase in parallel energy. Later on, $l_{o}$ increases with time and the parallel energy decreases. As particles transition from mirroring to transiting, $l_{o}$ converges to $L/4.0$ For some initial pitch angles slightly larger than the loss-cone angle limit, the transition from mirroring to transiting motion may occur at early times, while the flux surface is still contracting. If $l_{o}$ was increasing prior to the transition time, the parallel energy ratio switches from a loss to a gain (see, for example $\theta_{1i} = 60^\circ$ in Fig.\ \ref{fig:Island1_orbit_length}). If $l_{o}$ was decreasing prior to that time, the parallel energy ratio is always larger than $1$ (see, for example $\theta_{1i} = 25^\circ$ in Figure \ref{fig:Island2_orbit_length}).

To better distinguish between mirroring and transiting behaviors, the evolution of particle pitch angles on two flux surfaces is shown in panels b) of Figs.\ \ref{fig:Island1_orbit_length} and \ref{fig:Island2_orbit_length}. For flux surface 5 of Island 1, whose initial mirror ratio is $\approx 1.6$, particles with initial pitch angles below $\approx 52^\circ$ are inside the loss cone (see Appendix \ref{sec:appen_A}). For flux surface 8 of Island 2, whose initial mirror ratio is $\approx 6.15$, particles with initial pitch angles below $\approx 24^\circ$ are inside the loss cone. If pitch angles were initially isotropically distributed among the particles inside these islands, many more particles would experience mirroring in Island 2 than in Island 1. Pitch angles increase for those cases where the perpendicular energy gain is larger than the parallel energy gain. Pitch angles decrease at $t_{i}$ for Island 2 because there is a slight decrease in perpendicular energy as $B_{1}$ decreases at that time. 

The weighting coefficients of the energy gain components (first and second terms of Equation \ref{eq:en_gain_results}) depend on the particle's initial pitch angle. Both factors contribute to the total energy amplification occurring in the islands, but in different proportions for the mirroring and transiting particles. The larger the initial pitch angle, the stronger the perpendicular gain contribution to the total energy gain. On the other hand, the smaller the initial pitch angle, the larger the parallel contribution to the energy gain. The total perpendicular and parallel contributions to the energy gain are shown in the bottom panels of Figures \ref{fig:Island1_energy_ratios} and \ref{fig:Island2_energy_ratios}. The largest energy gains for Island 1 are due mostly to energy gain in the parallel direction and mostly for transiting particles. The opposite is true for Island 2: the largest energy gains are due mostly to energy gain in the perpendicular direction, for mostly mirroring particles. Consequently, panels a) and d) most closely resemble each other for Island 1, while panels a) and c) do so for Island 2.

\begin{figure}[h!]
 \centering
   \resizebox{6.0in}{!}{\includegraphics{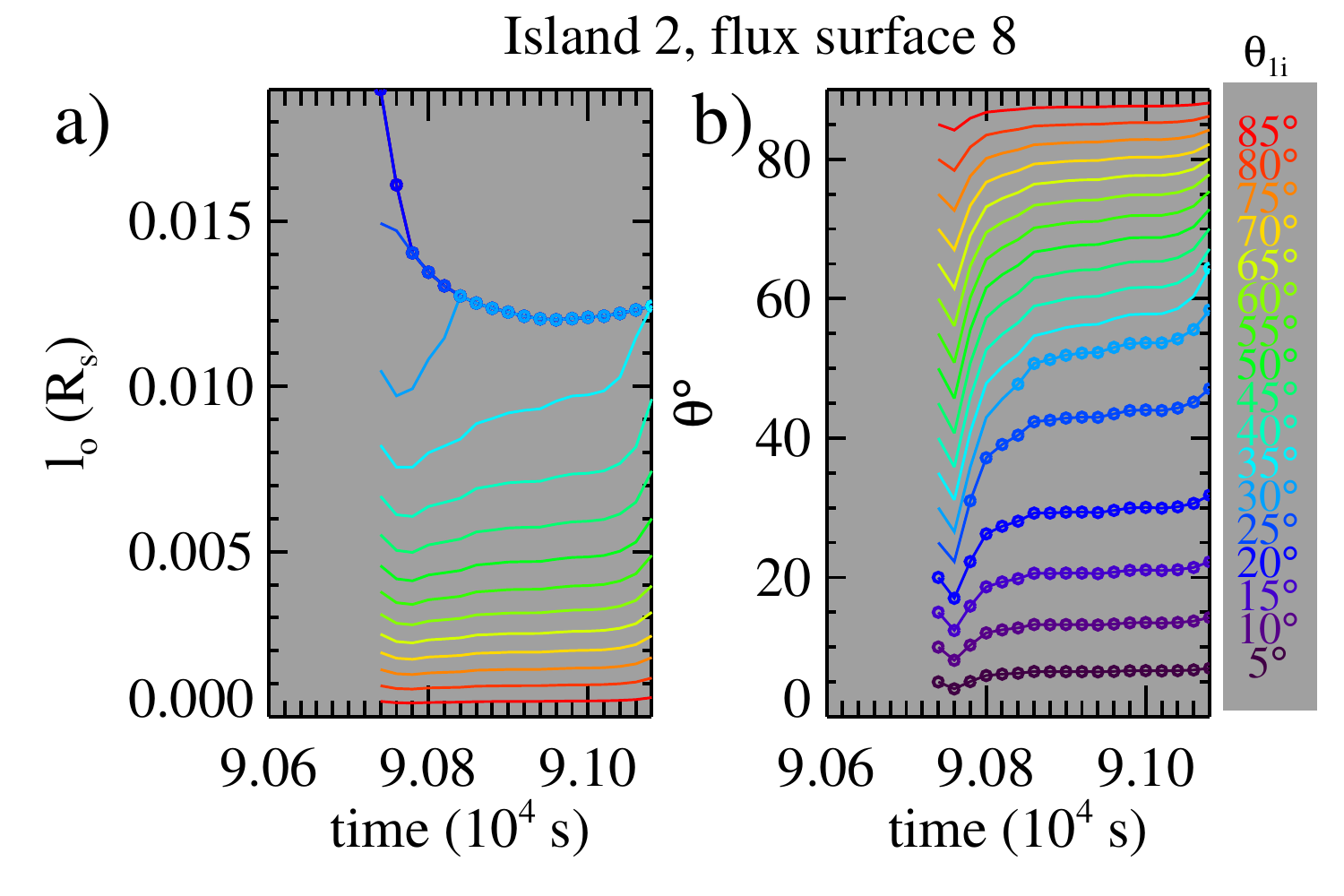}}
    \caption {Island 2, flux surface $8$. Evolution of $l_{o}$ and $\theta$ for the same color-coded initial pitch angles as in Figure \ref{fig:Island2_energy_ratios}. a) $l_{o}$. b) $\theta$. An animation showing $l_{o}$ and $\theta$ for all flux surfaces of Island 2 is included in the online journal (f14\_movie.mpg).}  
\label{fig:Island2_orbit_length}
\end{figure}

 \section{Discussion}
\label{sec:discussion}

We calculated analytically the energy gains of particles that are confined to large-scale individual 2.5D magnetic islands generated by spatially and temporally variable reconnection in a single CS. Due to contraction of the islands as they move along the sheet after being formed, their particles are subject to type A and type B \citet{Fermi_1949} acceleration. Our calculations extend an analysis by \citet{Drake_2006} of particle motion and adiabatic invariants in kinetic systems. We applied the method to an MHD-scale system including plasma compressibility and a finite guide field for consistency with the solar flare environment.  As derived in Appendix \ref{sec:appen_A}, a particle's energy gain depends upon its initial pitch angle and on the changes in length, magnetic field strength, and mirror ratio of the flux surface along which the particle moves. The energy gain is different in the directions parallel and perpendicular to the magnetic field, so that an initially isotropic distribution of particles becomes increasingly anisotropic over time.

To calculate the island formation and evolution leading to particle energy gains in flares, we performed a 2.5D simulation of a breakout CME/EF with the highest resolution to date (8 levels of refinement). Fast reconnection in our simulated flare CS is intermittent and forms large-scale magnetic islands. We analyzed in detail two sunward-moving islands formed by reconnection in the flare CS below the CME, from their birth to their arrival at the top of the flare arcade. These islands contract, their field strengths increase, and they have relatively large magnetic mirror ratios. We found that individual islands are candidates for only modest acceleration of particles intermittently in flares, yielding energy gain ratios of $1$--$5$. This energy gain is higher than those shown in Figure 2 of \citet{Drake_2006} and Figure 5 of \citet{Drake_2010} due to the compressibility of the plasma in our simulations. The mechanism should produce sporadic emission because island formation is intermittent. A large number of particles is accelerated while an island evolves, then streamed toward the footpoints or released into the top of the arcade by reconnection. When a new island is formed, a new set of particles is accelerated.

The modest energy ratios obtained on flux surfaces within single islands cannot boost superthermal electrons sufficiently to account for the observed flare HXR and microwave emissions. However, we expect that a larger number of islands would be formed in simulations with finer grids at lower resistivity. Higher resolution simulations are needed to test this in our global simulation. In this case, particles near the outermost flux surfaces could ``jump'' between islands at the X-null that separates one island from the next, as described by \citet{Drake_2006}. In this way, a particle needs to visit only a few islands to increase its energy by two orders of magnitude. It is currently impossible to test this hypothesis with even state-of-the-art PIC simulations (e.g., \citealt{Daughton_2009, Daughton_2014}), for the parameters of our simulation. To corroborate our analytical results, look for evidence of particle transfer from island to island, and examine the particle acceleration that occurs during fast island mergers at the top of the flare arcade, we are currently developing a ``test particle'' method. In this approach, the individual guiding-center orbits of an ensemble of test particles are calculated in the evolving electromagnetic field of the MHD simulation.

\citet{Drake_2006} report that the back pressure from accelerated particles limits the contraction achieved. If many particles are accelerated, large amounts of energy will be transferred to them, reducing the total available free magnetic energy. These nonlinear effects are not included in our, or any other, purely MHD simulation, because calculating the self-consistent simultaneous evolution of the particles and the electromagnetic field requires a fully kinetic approach.  A kinetic simulation also might reveal disruptive phenomena such as mirror and firehose instabilities, driven by the anisotropic particle distribution functions that develop due to the island contraction \citep{Drake_2010}. Furthermore, MHD shocks form in our flare current sheet (KAD12), and shock electric fields at kinetic scales also could contribute to particle acceleration. The only potentially feasible approach to solving the global problem of eruptive-flare evolution together with the local problem of self-consistent nonlinear feedback between particles and fields would seem to be an embedded PIC method \citep[e.g.,][]{Daldorff_2014}.

As shown for Island 1, island coalescence may increase the energy gain. Coalescence is favored in kinetic systems with many current layers, but there is no evidence (observational or numerical) for a flare CS to have multiple layers or to be folded upon itself. Reconnection in our flare CS is globally driven, as opposed to kinetic scenarios where reconnection is spontaneous or triggered by a small perturbation. Higher resolution may increase the coalescence frequency in our simulations. In more complicated 3D geometries, current filamentation may occur with plasmoids forming at multiple reconnection sites and interacting and merging. High-resolution 3D simulations capable of resolving a wide range of island sizes are needed to address these issues.

We show that our analysis is valid for a significant fraction of the coronal electron population over a reasonable range of preflare temperatures. The number of electrons in our simulated islands is large, of order $10^{35}$--$10^{36}$ per radian of angular extent along the PIL of the flare arcade. Thus, it is theoretically possible to accelerate a large number of electrons to higher energies by this macroscopic process. Our simulated active region is about an order of magnitude larger than a typical solar active region, however, which suggests that the sizes of the islands would be reduced by a similar factor in an actual flare. On the other hand, the electron density in our simulation is significantly lower than is typical of the flaring corona. These effects of length scale and plasma density would compensate each other, at least partially, in a more realistic configuration. These promising results need to be tested in a configuration with scales and densities characteristic of actual observed flares. As explained in KAD12, the timescale of our simulation is about $40$ times larger than that of a typical active-region flare. The lifetime of our islands (hundreds of seconds) corresponds to a few seconds of typical active-region flare times, comparable to the duration of some of the observed HXR spikes.

Most flare HXRs are produced by bremsstrahlung emitted by energetic electrons colliding with ambient charged particles. In principle, this process could be calculated from the results of our simulation, but it is beyond the scope of the present paper. The energetic-electron spectrum must be convolved with the cross sections for bremsstrahlung emission and the local plasma density to generate the HXR photon intensity \citep[e.g.,][]{Kontar_2011}. Within the confines of MHD, we plan to use a test particle approach to track the evolving energetic-electron spectrum. However, a comprehensive determination would require a self-consistent particle calculation including all sources of acceleration, transport, and interaction with the background plasma \citep[e.g.,][]{Liu_2009_I}.

Early islands, such as Island 1, are formed by reconnection of highly sheared field lines (large toroidal field). Its mirror ratios are small and most of the particles orbiting this island are transiting along the helical field lines. Transiting particles have the highest energy gains, mostly in the parallel direction. In 3D, particles on these field lines would precipitate toward their footpoints near the PIL. If these conditions are typical of islands formed in the early, impulsive phase of flares, then this could help explain the observed HXR footpoint sources observed in a flare.  

On the other hand, islands formed later in the flare, such as Island 2, have less sheared field. The large mirror ratios yield a population of predominantly mirroring particles, which are trapped above the top of the flare arcade until the island collides with the top of the arcade. After the collision with the arcade, these accelerated particles may find a thick target and emit in HXR (\citealt{Lee_2013}), explaining the observed flare loop-top sources. Note that a small fraction of particles in Island 1 are also trapped above top of the arcade and some in Island 2 are transiting. Therefore some emission might be seen at the footpoints and loop-top sources simultaneously. We have not analyzed the merger of the islands with the top of arcade, which could be a subsequent acceleration mechanism for those trapped particles (e.g., \citealt{Sakai_1992, Kolomanski_2007, Milligan_2010, Karlicky_2011, Nishizuka_2013}).

 \acknowledgements

We thank Prof. Jim Drake, Dr. Spiro Antiochos, and Dr. Peter Wyper for helpful comments and advice. This research was supported in part by the NASA Heliophysics Supporting Research program. Resources supporting this work were provided by the NASA High-End Computing Program through the NASA Center for Climate Simulation at Goddard Space Flight Center. 

\begin{appendices}

\numberwithin{equation}{section}

\section{Analytical Solutions for Mirroring and Transiting Particles}
\label{sec:appen_A}

We calculate in detail the energy gained by a particle orbiting along a magnetic flux surface in a 3D flux rope. A cross section of such a flux rope is shown schematically as a 2D island in Figure \ref{fig:island_cartoon}. The two principal assumptions underlying our approach are: (1) adiabatic invariance, i.e., the characteristic time over which the magnetic field changes is much longer than the time required for the particle to complete a periodic orbit; and (2) collisionless motion, i.e., the characteristic time over which the particle suffers a significant deflection of its motion by collisions is much longer than the orbit time. The validity of these assumptions in the flaring corona is tested in Appendix \ref{sec:appen_C} below.

The adiabatic invariants arise from conservation relations for the action $A$ associated with Hamiltonian coordinates of the motion, $q$, and their conjugate momenta, $p$, over a complete orbit of periodic motion, 
\begin{equation}
     A = \oint p dq
\end{equation}
\citep[e.g.,][]{Kruskal_1962, Kulsrud_2005}. For a charged particle moving in a magnetic field, the first invariant is $\mu$, the magnetic moment, associated with the gyromotion of the particle perpendicular to the field direction. The second invariant is $J$, the parallel action, associated with the longitudinal motion of the particle parallel to the field direction. They are expressed by 
\begin{align}
     \label{eq:microconsm}
     \mu &= \frac{V_{\perp}^2}{B},\\
     \label{eq:microconsj}
     J &  = \oint V_{\parallel} \,dl,
\end{align}
where $V_{\perp}$ and $V_{\parallel}$ are the particle velocity components perpendicular and parallel to the magnetic field, respectively, $B$ is the field strength, and $l$ is the arc-length coordinate along the field line. The longitudinal motion is periodic if either (1) the particle is trapped along a finite length of the flux rope due to magnetic mirroring at both ends of its orbit, or (2) the flux rope structure is periodic along its length, as is true in our simple simulation setup with axial symmetry along the out-of-plane direction in Figure \ref{fig:island_cartoon}. The invariance of $\mu$ follows simply from conservation of the axial magnetic flux passing through the particle's gyro-orbit; the invariance of $J$ is less intuitive, and demonstrating it is rather intricate mathematically \citep{Northrop_1960}.

Conservation of the magnetic moment implies that the perpendicular speed $\left\vert V_{\perp} \right\vert$ varies with the magnetic field strength $B(l)$ along the field line. Therefore, the parallel speed $\left\vert V_{\parallel} \right\vert$ does, as well. To make progress, we adopt the following model profile for the magnetic field strength, 
\begin{equation}
   \label{eq:B_prof}
    B = B_{1} + \left( B_{2}-B_{1} \right) \sin^{2} \left( 2 \pi \frac{l}{L} \right).
\end{equation}
$B_{1}$ and $B_{2}$ are the minimum and maximum field strengths, respectively, along the flux surface, and $L$ is the length of one full turn of the flux rope. The resulting island is symmetric both left/right and up/down, as sketched in Figure \ref{fig:island_cartoon}, possessing two equal minima $B = B_{1}$ at $l = 0, L/2$ and two equal maxima $B = B_{2}$ at $l = L/4, 3L/4$. All three of the flux surface parameters are time dependent, in general.

Introducing the total speed, $V$, and the pitch angle, $\theta$, of the particle, we can write
\begin{align}
     \label{eq:vper}
     V_{\perp} &  = V \sin \theta, \\
          \label{eq:vpar}
     V_{\parallel} &= V \cos \theta.
\end{align}
Let $\theta_{1}$ be the pitch angle where the magnetic field strength is smallest, $B = B_{1}$. Conservation of the magnetic moment, $\mu$, and kinetic energy (i.e., $V$) along the orbit implies 
\begin{equation}
    \label{eq:mu_const}
    \mu = \frac{V^{2} \sin^{2}\theta}{B} = \frac{V^{2} \sin^{2}\theta_{1}}{B_{1}} = \hbox{constant}
\end{equation}
after using Equations (\ref{eq:microconsm}) and (\ref{eq:vper}). From this expression, we derive 
\begin{align}
 \label{eq:sin_theta}
    \sin^{2}\theta &= \frac{B}{B_{1}} \sin^{2}\theta_{1} \nonumber \\
                   &= \sin^{2}\theta_{1} + \left( \frac{B}{B_{1}} - 1 \right) \sin^{2} \theta_{1} \\
                   &= \sin^{2}\theta_{1} + \left( \frac{B_{2}}{B_{1}} - 1 \right) \sin^{2} \theta_{1} \sin^{2} \left( 2 \pi \frac{l}{L} \right) \nonumber
\end{align}
after using the model profile in Equation (\ref{eq:B_prof}). For the parallel velocity, we then find 
\begin{align}
 \label{eq:v_par}
   V_{\parallel} &= V \left[ 1 - \sin^{2}\theta \right]^{1/2} \nonumber \\
                 &= V \left[ 1 - \sin^{2}\theta_{1} - \left( \frac{B_{2}}{B_{1}} - 1 \right) \sin^{2}\theta_{1} \sin^{2} \left( 2 \pi \frac{l}{L} \right) \right]^{1/2} \\  
  	         &= V \cos \theta_{1} \left[1 - \left( \frac{B_{2}}{B_{1}} - 1 \right) \tan^{2}\theta_{1} \sin^{2} \left( 2 \pi \frac{l}{L} \right) \right]^{1/2}. \nonumber 
\end{align}
It is convenient to introduce the new parameter $\beta$, 
\begin{equation}
    \label{eq:beta}
    \beta \equiv \left( \frac{B_{2}}{B_{1}} - 1\right)^{1/2} \tan\theta_{1}.
\end{equation}
If $\beta < 1$, i.e., if the particle's pitch angle $\theta_{1}$ at $B = B_{1}$ is sufficiently small, then $V_{\parallel}$ in Equation (\ref{eq:v_par}) is never zero; this particle is {\it transiting} and samples the entire flux surface. On the other hand, if $\beta \geq 1$, i.e., the pitch angle $\theta_{1}$ is sufficiently large, then $V_{\parallel}$ goes to zero at two positions within a distance $l_r < L/4$ of $l = 0$ or $l = L/2$ (the particle can be trapped on either side of the island); this particle is {\it mirroring} and samples only a subset of the flux surface. The following condition is satisfied at the mirror points:
\begin{equation}
    \beta^2 \sin^2 \left( 2 \pi \frac{l_{r}}{L} \right) = 1. 
     \label{eq:orbit_length}
\end{equation}
The condition on $\beta$ is equivalent to the well-known definition of the loss cone: a particle mirrors if $\sin\theta_{1} \geq \sqrt{B_{1}/B_{2}} = 1/\sqrt{R_{m}}$, where $R_{m} = B_{2}/B_{1}$ is the mirror ratio, and transits (is lost) if $\sin\theta_{1} < 1/\sqrt{R_{m}}$.

For either type of orbit, the parallel action integral in Equation (\ref{eq:microconsj}) reduces to the form 
\begin{equation}
       J = 4 V_{\parallel 1} \int_{0}^{l_{o}} \left[1 -  \beta^2 \sin^{2} \left( 2 \pi \frac{l}{L} \right)\right]^{1/2} dl
\end{equation}
after substituting the expression in Equation (\ref{eq:v_par}) and exploiting the four-fold symmetry of the orbit between $l = 0$ and $l = l_{o}$. The upper limit of the integral is $l_{o} = L/4$ for a transiting particle, and $l_{o} = l_{r}$ for a mirroring particle. In the first case, for $\beta < 1$, a variable change yields 
\begin{equation}
       J = \frac{2}{\pi} L V_{\parallel 1} \int_{0}^{\pi/2} \left[1 - \beta^2 \sin^{2} t \right]^{1/2} dt;
\end{equation}
in the second case, for $\beta \ge 1$, a different variable change gives 
\begin{equation}
       J = \frac{2}{\pi} L V_{\parallel 1} \beta^{-1} \int_{0}^{\pi/2} \frac{\cos^{2} t}{\left[1 - \beta^{-2} \sin^{2} t \right]^{1/2}} dt. 
\end{equation}
The above expressions for $J$ are monotonic over the entire range of $\beta$ and smoothly continuous at $\beta = 1$. Introducing the function $\mathcal{F}(\beta)$, we write $J$ as
\begin{equation}
   \label{eq:f_action}
       J = L V_{\parallel 1} \mathcal{F}(\beta).
\end{equation}
The preceding integrals can be expressed in closed form (see \citealt[integrals 3.671.2 and 3.671.1]{Gradshteyn_2007}),
\begin{equation}
   \label{eq:trans_vs_mirr}
       \mathcal{F}(\beta) = \left\{  
                        \begin{array}{lr}
                                F(\frac{1}{2},-\frac{1}{2};1, \beta^{2}), & 0 \leq \beta < 1\hbox{  (transiting);}  \\
                                  & \\
                                  (2\beta)^{-1} F(\frac{1}{2},\frac{1}{2};2, \beta^{-2}), & \beta \geq 1 \hbox{  (mirroring),} 
                                  \\
                         \end{array}
                          \right.
\end{equation}
where $F(a,b,c,z)$ is Gauss's hypergeometric function.

We now derive a conserved expression involving $V_{\parallel}$ from the definition of the magnetic moment,
\begin{equation}
\label{eq:mu_const2}
       \frac{V_{\parallel} \tan \theta}{\sqrt{B}} = \frac{V_{\perp}}{\sqrt{B}} = \sqrt{\mu} = \hbox{constant}.
\end{equation}
Evaluating this expression at position ``1'', substituting the result into Equation (\ref{eq:f_action}), and using Equation (\ref{eq:trans_vs_mirr}) leads to 
\begin{equation}
       J = \sqrt{\mu} \frac{L \sqrt{B_{1}}}{\tan \theta_{1}} \mathcal{F}(\beta) = \sqrt{\mu} \frac{L \sqrt{B_{1}}}{\tan \theta_{1}} \mathcal{F} \left( \left[ \frac{B_{2}}{B_{1}} - 1\right]^{1/2} \tan\theta_{1} \right) = \hbox{constant}.
\end{equation}
Finally, using the subscript ``i'' to denote the initial values of all quantities and remembering that $\mu$ is conserved, we obtain 
\begin{equation}
     \label{eq:transcendental}
        \frac{L \sqrt{B_{1}}}{\tan \theta_{1}} \mathcal{F} \left( \left[ \frac{B_{2}}{B_{1}} - 1\right]^{1/2} \tan\theta_{1}\right) = \frac{L_{i} \sqrt{B_{1i}}}{\tan \theta_{1i}} \mathcal{F} \left( \left[ \frac{B_{2i}}{B_{1i}} - 1\right]^{1/2} \tan\theta_{1i}\right).
\end{equation}
This is a transcendental equation in $\tan \theta_{1}$ that can be solved numerically. The new pitch angle $\theta_{1} $ of any particle is then a function of its initial pitch angle and of the initial and new properties of the flux surface: the field strength, mirror ratio, and length of one turn of a field line. Whether the particle is transiting or mirroring is determined by the new values of $\theta_{1}$ and the mirror ratio $B_{2}/B_{1}$.

We now calculate the energy gain of a particle $\mathcal{E}$, the ratio of its new kinetic energy to its initial value, evaluated at the minimum of the field, as follows:
  \begin{align}
      \label{eq:en_gain}
      \mathcal{E}  &= \frac{V^2_{\perp 1} + V^2_{\parallel 1}}{V^2_{\perp 1i} + V^2_{\parallel 1i}} \nonumber \\
                   &= \left( \frac{V^2_{\perp 1i}}{V^2_{\perp 1i} + V^2_{\parallel 1i}} \right) \mathcal{E}_{\perp} + \left( \frac{V^2_{\parallel 1i}}{V^2_{\perp 1i} + V^2_{\parallel 1i}} \right) \mathcal{E}_{\parallel} \\
                   &= \sin^2 \theta_{1i} \hbox{ } \mathcal{E}_{\perp} + \cos^2 \theta_{1i} \hbox{ } \mathcal{E}_{\parallel}. \nonumber
  \end{align}
$\mathcal{E}_{\perp}$ and $\mathcal{E}_{\parallel}$ are the energy gains in the directions perpendicular and parallel to the magnetic field, respectively. From Equations (\ref{eq:mu_const}) and ( \ref{eq:mu_const2}),
 \begin{equation}
      \label{eq:E_perp}
      \mathcal{E}_{\perp} = \frac{V^2_{\perp 1}}{V^2_{\perp 1i}} = \frac{B_{1}}{B_{1i}},
 \end{equation}
  \begin{equation}
      \label{eq:E_paral}
      \mathcal{E}_{\parallel} = \frac{V^2_{\parallel 1}}{V^2_{\parallel 1i}} = \frac{B_{1}}{B_{1i}} \frac{\tan^2 \theta_{1i}}{\tan^2 \theta_{1}}.
 \end{equation}
The above energy factors show that initially isotropic energies become increasingly anisotropic as islands evolve. In all cases, the perpendicular energy increases if the minimum field strength increases. The parallel energy changes in a complicated way depending upon the initial pitch angle of the particle and the three parameters describing the magnetic field profile.

There are two useful limiting behaviors for the parallel energy gain: 1)  Uniform field ($B_{1} =B_{2}$, $\beta \rightarrow 0$); and 2) Strong magnetic trap ($B_{2} >> B_{1}$, $\beta \rightarrow \infty$). For these cases, from Equation \ref{eq:transcendental} and considering the fact that $F(a,b,c,0) = 1$, the parallel energy gain takes the following forms 
  \begin{align}
      \label{eq:par_large_beta}
      \mathcal{E}_{\parallel} = 
         \left\{  
             \begin{array}{lr}
                   \left( L_{i} / L \right)^2, &  B_{1} = B_{2}, B_{1i} = B_{2i};   \\ 
                    & \\
                  \left( L_{i} / L \right)  \left( B_{2} / B_{2i} \right)^{1/2}, &  B_{2} >> B_{1}, B_{2i} >> B_{1i}.
       \end{array}
                          \right.
  \end{align}
For the uniform field case, the parallel energy increases if the island contracts, but is not affected by a change in field strength. For the strong magnetic trap, the parallel energy could also increase if the island contracts, as long as the field satisfies that $B_{2}/B_{2i} > L^2/L^2_{i}$. 

If the initial $\theta_{i}$ of a particle is equal to zero, then $\mu=0$ and $\theta = 0$ for all times. This purely parallel motion has an increase in energy $\mathcal{E} = \mathcal{E}_{\parallel} = (L_{i}/L)^{2}$. Another special case occurs when $\theta_{1} = \pi/2$, which corresponds to pure gyromotion at the minimum of the field, with no parallel displacement (different from $\theta =  \pi/2$ at a mirroring point). In this case, $\mathcal{E} = \mathcal{E}_{\perp} = B_{1}/B_{1i}$.
 
\section{Algebraic Expressions for Uniform Magnetic Field Strength}
\label{sec:appen_B}

We now specialize the calculations of Appendix \ref{sec:appen_A} to derive algebraic expressions for the energy gain of particles traversing a flux surface with uniform field strength. In this special case, all particles are transiting, because the mirror ratio is unity. The energy gain can be expressed simply as functions of the macroscopic quantities: mass density, island length, and magnetic field strength.

Consider a flux region, such as those shaded red in Figure \ref{fig:island_cartoon}ab. We assume that its width $w$ is small and the magnetic field is uniform inside the flux surface. The field components are poloidal $B_{p}$ (in-plane) and toroidal $B_{t}$ (out-of-plane). Thus, $B=B_{1}=\sqrt{B_t^2+B_p^2}$. If the 2D projected length of the flux surface is $L_{p}$, the length of one turn of the field lines on the flux surface is 
\begin{equation}
     L = {\left( 1 + \frac{B_t^2}{B_p^2} \right)^{\frac{1}{2}}} L_{p}.
\end{equation}
There are three macroscopic constraints for particles orbiting inside this flux region: conservation of mass $M$, toroidal flux $\Psi_{t}$, and poloidal  flux $\Psi_{p}$:
\begin{align}
     \label{eq:macroconsm}
     M  \approx  n w L_{p} & = n_{i} w_{i} L_{p,i},\\
     \label{eq:macroconst}
     \Psi_t \approx  B_{t} w L_{p} & = B_{t,i} w_{i} L_{p,i},\\
     \label{eq:macroconsp}
     \Psi_p  \approx  B_{p} w & = B_{p,i} w_{i},
\end{align}
where $n$ is the plasma density. By combining Equations (\ref{eq:macroconsm}), (\ref{eq:macroconst}), and (\ref{eq:macroconsp}) we find that 
\begin{align}
     \label{eq:btoro2bpolo}
     \frac{B_t}{B_p} &= \frac{B_{t,i}}{B_{p,i}} \frac{L_{p,i}}{L_{p}},\\
     \label{eq:bpoloratio}
     \frac{B_p}{B_{p,i}} &= \frac{w_i}{w} = \frac{n}{n_i} \frac{L_{p}}{L_{p,i}}.
\end{align}
After some simple algebra, Equation (\ref{eq:E_perp}) and the top line of Equation (\ref{eq:par_large_beta}) can be expressed as
\begin{equation}
     \label{eq:eperp}
     {\mathcal{E}_{\perp}} = \frac{n}{n_i} \left( \frac{L_{p}^2}{L_{p,i}^2} + \frac{B_{t,i}^2}{B_{p,i}^2} \right)^{\frac{1}{2}} \Big/ \left( 1 + \frac{B_{t,i}^2}{B_{p,i}^2} \right)^{\frac{1}{2}},
\end{equation}
{and} 
\begin{equation}
     \label{eq:eparallel}
     {\mathcal{E}_\parallel} =  \left( 1 + \frac{B_{t,i}^2}{B_{p,i}^2} \right) \Big/ \left( \frac{L_{p}^2}{L_{p,i}^2} + \frac{B_{t,i}^2}{B_{p,i}^2} \right).
\end{equation}
For particles having isotropic initial velocities, $V_{{\perp}i1}^2 = 2V_{{\parallel}i1}^2$ ($\tan \theta_{1i} =\sqrt{2} $), Equation \ref{eq:en_gain} can expressed as 
\begin{equation}
     \label{eq:eisotropic}
     {\mathcal{E} = \frac{2}{3} \mathcal{E}_\perp + \frac{1}{3} \mathcal{E}_\parallel.}
\end{equation}
Assuming that $B_{t,i} = 0$ simplifies Equations (\ref{eq:eperp}) and (\ref{eq:eparallel}) greatly, and Equation (\ref{eq:eisotropic}) becomes
\begin{equation}
     \label{eq:eguidoni}
     {\mathcal{E}} \rightarrow {\frac{2}{3} \frac{n}{n_i} \frac{L_{p}}{L_{p,i}} + \frac{1}{3} \frac{L_{p,i}^2}{L_{p}^2}}.
\end{equation}
Assuming also that the flow is incompressible leads to 
\begin{equation}
     \label{eq:edrake}
     {\mathcal{E}} \rightarrow {\frac{2}{3} \frac{L_{p}}{L_{p,i}} + \frac{1}{3} \frac{L_{p,i}^2}{L_{p}^2}}.
\end{equation}
The last expression was derived and discussed by \citet{Drake_2010}.

The energy multiplication factors above show that initially isotropic particle distributions become increasingly anisotropic as the island perimeter shrinks. Indeed, in the zero guide field, incompressible limit 
of (\ref{eq:edrake}), the perpendicular energy (first term) actually decreases, but the parallel energy (second term) increases even faster, so that overall the particle accrues energy. The decrease in perpendicular energy occurs because, as the island perimeter $L_{p}$ shrinks, the width $w$ increases to preserve the enclosed volume, $B_p$ decreases to conserve the enclosed poloidal flux, and $V_\perp$ decreases to keep the magnetic moment invariant, 
\begin{equation}
     \label{eq:incompressible1}
     \frac{L_{p}}{L_{p,i}} = \frac{w_i}{w} = \frac{B_p}{B_{p,i}} = \frac{V^2_{\perp}}{V^2_{{\perp}i}},
\end{equation}
when $n = n_i$, $B_t = 0$. The increase in parallel energy maintains the constancy of the parallel action as the perimeter shrinks,
\begin{equation}
     \label{eq:incompressible2}
     \frac{L_{p}}{L_{p,i}} = \frac{V_{{\parallel}i}}{V_{\parallel}},
\end{equation}
and an increase in the total particle energy results, as described by \citet{Fermi_1949}. Allowing for plasma compression ($n \ne n_i$), as in (\ref{eq:eguidoni}), changes the outcome qualitatively if the width $w$ decreases, rather than increases, as the perimeter $L_{p}$ shrinks. This requires a sufficiently large density enhancement, 
\begin{equation}
     \label{eq:compressible1}
     \frac{n}{n_i} > \frac{L_{p,i}}{L_{p}},
\end{equation}
in which case 
\begin{equation}
     \label{eq:compressible2}
     \frac{V^2_{\perp}}{V^2_{{\perp}i}} = \frac{B_p}{B_{p,i}} = \frac{w_i}{w} = \frac{L_{p}}{L_{p,i}} \frac{n}{n_i} > 1.
\end{equation}
In this regime, both the perpendicular and parallel energies increase as the island contracts. At smaller compressions, the perpendicular energy decreases as before, but more slowly than in the incompressible limit, so that overall the particle energy gain is greater. A finite out-of-plane component of the magnetic field ($B_t \ne 0$) has the net effect of moderating the impact of changes in the island perimeter: the parallel energy increases less rapidly, and the perpendicular energy factor that depends upon $L_{p}$ decreases less rapidly. Because a finite $B_t$ also is expected to reduce the compression $n / n_i$, however, the net effect seems to be a smaller change in the perpendicular energy, whether the latter increases or decreases.

\section{Conditions for Adiabatic Invariance}
\label{sec:appen_C}

We now estimate the range of validity of our analytical analysis presented in Appendix \ref{sec:appen_A}. A major assumption of those calculations is the adiabatic invariance of the magnetic moment and parallel action of the accelerated particle. This requires, first, that the associated particle orbit (gyromotion, mirroring, or transiting) be completed much more quickly than the magnetic field changes in time. The characteristic speed for magnetic field changes in the islands is the Alfv\'en speed, $V_A$. $V_A$ is large compared to the proton (or other ion) thermal speed in the CS of a flare, where the plasma $\beta$ is small, so the analysis above is appropriate only for protons and ions that already are strongly superthermal. This is not necessarily the case for electrons, however, whose speed $V_e$ at energy $E_e$ is
\begin{equation}
     \label{eq:ve}
     V_e = \left( \frac{2E_e}{m_e} \right)^{\frac{1}{2}} \approx 5000 \left( \frac{E_e}{E_6} \right)^{\frac{1}{2}}~{\rm km~s}^{-1},
\end{equation}
where $E_6$ is the thermal energy corresponding to a coronal electron temperature $T_e = 10^6~{\rm K}$. The normalization speed, 5000 km s$^{-1}$, is comparable to Alfv\'en speeds estimated in the corona (e.g., at field strength $B = 10~{\rm G}$ and number density $n = 10^9~{\rm cm}^{-3}$). Thus, for moderately superthermal electrons at $1 \times 10^6~{\rm K}$ and for thermal electrons at $10^7~{\rm K}$ -- a typical range of preflare temperatures -- the adiabatic invariance of the parallel action is a reasonable approximation. A second condition also applies, however: the particle must not suffer collisions while executing its trajectory. The mean free path $D_e$ is
\begin{equation}
     \label{eq:de}
     D_e = \frac{3}{4 \sqrt{2\pi} n \lambda e^4} E_e^2 \approx 100 \left( \frac{10^9}{n} \right) \left( \frac{10}{\lambda} \right) \left( \frac{E_e}{E_6} \right)^2~{\rm km},
\end{equation}
where $\lambda$ is the Coulomb logarithm \citep{Spitzer_1962}. The normalization length, 100 km, is small compared to the size of flaring regions, but the islands are small structures within the flare CS. In addition, $D_e$ varies quadratically with the particle energy, so the mean free path rises rapidly for moderately superthermal electrons at ambient coronal temperatures, and it is very large for thermal electrons at enhanced preflare temperatures. Thus, again we conclude that the adiabatic invariance of the parallel action is a reasonable approximation. For at least a significant fraction of the coronal electron population, therefore, our analysis is adequate for a first assessment of the contribution of magnetic-island contraction to particle acceleration in solar flares. In principle, all particles are accelerated by the mechanisms described in this paper, but the energy gain of those that do not satisfy the above restrictions cannot be estimated by our equations based on adiabatic invariance.

\end{appendices}



\end{document}